\documentclass[12pt]{article}

\usepackage[superscript,sort]{cite}

\usepackage[pdftex]{graphicx}
\usepackage[usenames,pdftex]{color}
\definecolor{Red}{rgb}{1.0,0.0,0.0}

\usepackage[toc,page]{appendix}

\usepackage{amsthm,amscd,amsxtra,amsfonts,amsmath,amssymb,multirow}
\usepackage{wrapfig}
\usepackage[footnotesize]{caption}
\usepackage[tiny,compact]{titlesec}
\usepackage{ctable}
\usepackage{helvet}

\usepackage{subfigure}
\usepackage{tikz}
\usetikzlibrary{shapes,arrows}

\titlespacing*{\section}{0pt}{*0}{*0}
\titlespacing*{\subsection}{0pt}{*0}{*0}
\titlespacing*{\subsubsection}{0pt}{*0}{*0}
\titlespacing{\paragraph}{0pt}{*0}{*1}

\definecolor{MyPurple}{rgb}{1,0,1}

\newcommand{\beq}[1]{\begin{equation} \label{#1}}
\newcommand{\eeq}{\end{equation}}
\newcommand{\barray}{\begin{array}{ll}}
\newcommand{\earray}{\end{array}}

\begin{document}

\pagenumbering{roman}

\clearpage \pagebreak \setcounter{page}{1}
\renewcommand{\thepage}{{\arabic{page}}}

\title{Protein folding tames chaos
}

\author{
Kelin Xia$^1$
Guo-Wei Wei$^{1,2,3}$ \footnote{ Address correspondences  to Guo-Wei Wei. E-mail:wei@math.msu.edu}\\
$^1$ Department of Mathematics \\
Michigan State University, MI 48824, USA\\
$^2$ Department of Electrical and Computer Engineering \\
Michigan State University, MI 48824, USA \\
$^3$ Department of Biochemistry and Molecular Biology\\
Michigan State University, MI 48824, USA \\
}

\date{\today}
\maketitle

\begin{abstract}

Protein folding produces characteristic and functional three-dimensional structures from unfolded polypeptides or disordered coils. The emergence of extraordinary complexity in  the protein folding process poses astonishing challenges to theoretical modeling and computer simulations. The present work   introduces  molecular nonlinear dynamics (MND), or molecular chaotic dynamics, as a theoretical framework for describing and analyzing protein folding. 
We unveil  the existence of intrinsically low dimensional manifolds (ILDMs) in the chaotic dynamics of folded   proteins.  Additionally, we reveal that the transition from disordered to ordered conformations in protein folding increases the transverse stability of the ILDM. Stated differently, protein folding reduces the chaoticity of the nonlinear dynamical system, and  a folded protein has the best ability to tame chaos.  Additionally, we bring to light  the connection between the ILDM stability and the thermodynamic stability, which enables us to quantify the disorderliness and relative energies of folded, misfolded and unfolded protein states.  Finally, we exploit chaos for protein flexibility analysis and  develop a robust chaotic algorithm for the prediction of Debye - Waller factors, or temperature factors, of protein structures.
\end{abstract}
\maketitle


Anfinsen's dogma of sequence-structure-function \cite{Anfinsen:1973}, in which a protein's function depends on its uniquely folded three-dimensional (3D) structure and its structure is determined by  the amino acid sequence, is challenged due to the discovery that many partially folded or intrinsically unstructured proteins remain functional despite of the lack of  uniquely folded 3D structures \cite{Onuchic:1997,White:1999,Schroder:2005,Chiti:2006}. Kinetically and thermodynamically regulated competing pathways, including disordered aggregation, degradation, folding and unfolding, convert  linear chains of amino acids  translated from   sequences of mRNA into degraded fragments,  protofibrils, amyliod-like fribrils, amyliods, intrinsically disordered proteins, partially disordered proteins, and  folded structures \cite{Fischer:1894,NGo:1983,Schroder:2005,Chiti:2006,Uversky:2008}. The formation of disordered proteins is often exploited by living systems to perform novel and diverse biological functions. Unfortunately, aggregated or misfolded proteins are often associated with sporadic neurodegenerative diseases, such as mad cow disease, Alzheimer's disease and Parkinson's disease. Currently, there is a lack of efficient means for the characterization of  disordered aggregation and the quantification of  orderliness, which are crucial to the understanding of the molecular mechanism  of degenerative diseases.

\begin{figure}
\begin{center}
\begin{tabular}{cc}
\includegraphics[width=0.48\textwidth]{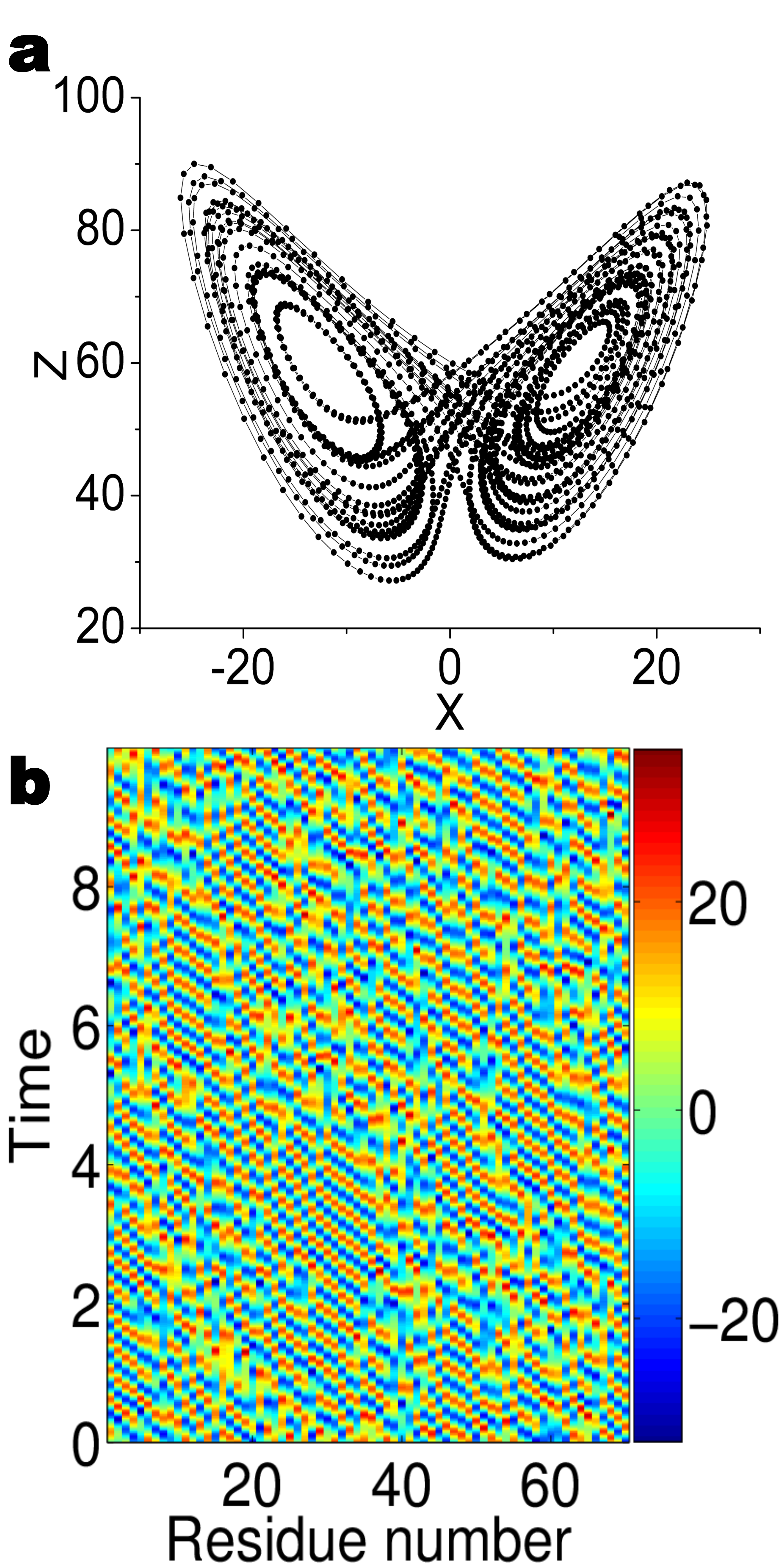}&
\includegraphics[width=0.48\textwidth]{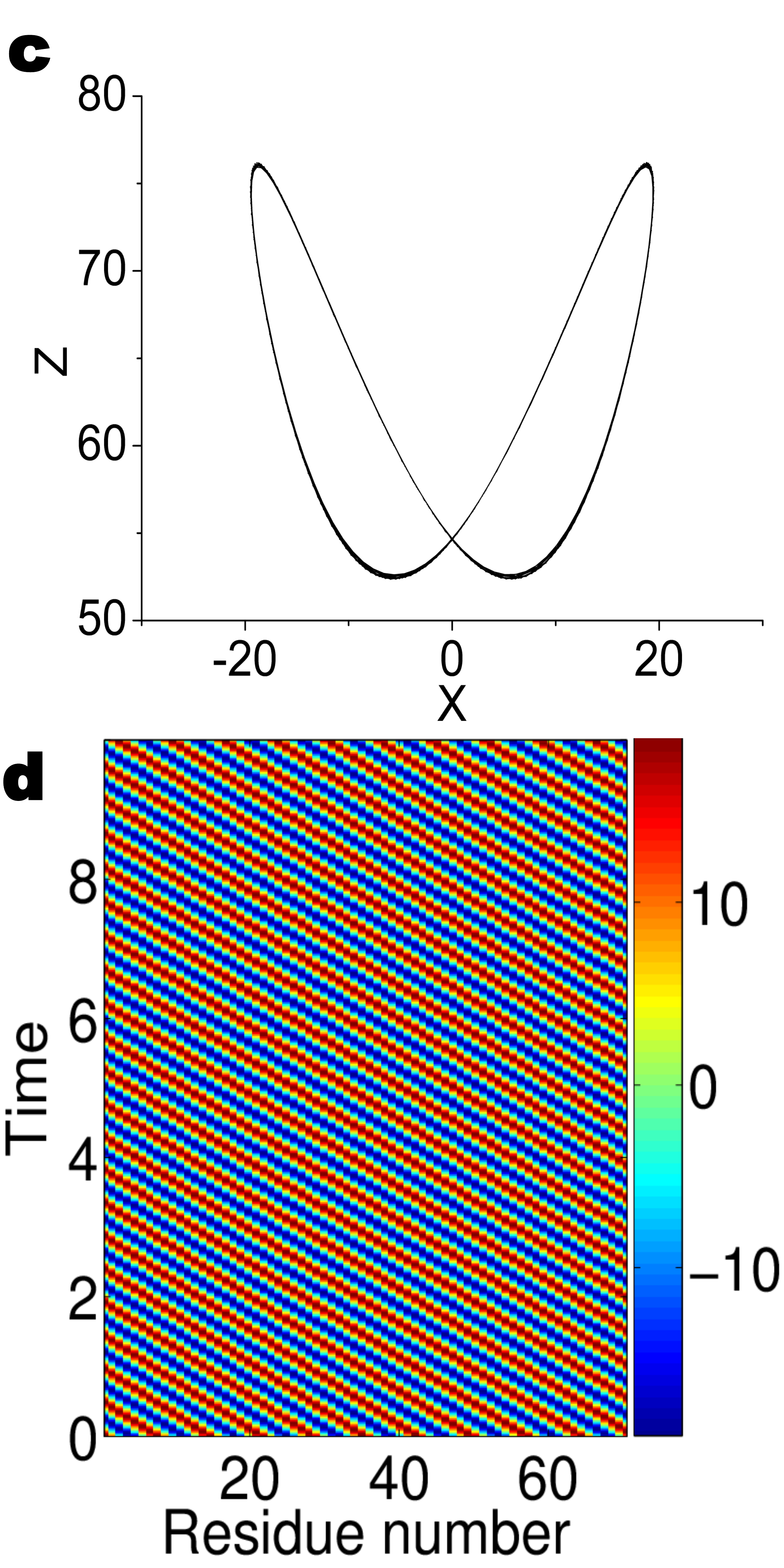}
\end{tabular}
\end{center}
\caption{Transition from chaos to periodicity in the   chaotic dynamics model (CDM) of bacteriocin AS-48 (PDB ID 1e68).
{\bf a} The butterfly wing pattern for one of 70 chaotic oscillators.
{\bf b} The solution of original 70 chaotic oscillators.
{\bf c} The periodic orbit of the ILDM for bacteriocin AS-48.
{\bf d} Bacteriocin AS-48 induced Hopf bifurcation from chaos. All of 70 nonlinear oscillators are in one lag synchronized periodic orbit.
}
\label{fig1}
\end{figure}

The emergence of complexity in self-organizing living systems, including protein folding, poses fabulous challenges to their quantitative description and prediction \cite{Bryngelson:1995,Dill:1997}.  Cyrus Levinthal suggested that  there are  near $10^{95}$  possible conformations for a relatively small polypeptide of 100 residues \cite{Levinthal:1969}, while an average human protein of 480 residues might have an astronomical number of  conformations. The complexity is extraordinary since human proteins are coded by over 20,000 genes. The straightforward sampling of the full  conformational space becomes unfeasible for  large proteins. It takes many months for molecular dynamics (MD) simulations, the main workhorse of computational protein folding, to come up with a very poor copy of what Nature administers perfectly within a tiny fraction of a second. In fact, disordered aggregation, unfolding and folding often occur at slower time scales and involve larger length scales, which are essentially intractable to full atomic simulations \cite{Bryngelson:1995,Dill:1997}. Coarse-grained (CG) representations of polypeptides are employed to  reduce  the number of degrees of freedom, extend molecular modeling,  and bridge with experimental observations. An active research topic is how to improve the accuracy of CG models so as to differentiate near degenerate energy landscapes of some conformations generated by the protein folding process. Elastic network models (ENMs), including Gaussian network model (GNM)  and anisotropic network model,  represent folded proteins as    elastic mass-spring networks to investigate its mechanical flexibility and long-time stability beyond the reach of molecular dynamics  \cite{Tasumi:1982,Brooks:1983,Levitt:1985,Tirion:1996,Bahar:1997}. In general,  ENMs can be viewed as a time-independent molecular mechanics derived from their corresponding time dependent molecular mechanics by using the time-harmonic approximation.
An underlying assumption adopted in all of the above-mentioned theoretical models is that protein folding, misfolding and aggregation are to be modeled with some deterministic  dynamical systems, which reinforce Anfinsen's dogma and exclude any unpredictability  and molecular degradation. However, the existence of  degraded fragments,  protofibrils, amyliod-like fribrils, amyliods, misfolds, and intrinsically disordered proteins highlights the fundamental limitation of current simulation models.

Chaos is ubiquitous in nature. The discovery of the sensitivity of initial conditions, one of three signatures of chaos,  dated to the 1880s by Henri Poincar\'{e} \cite{Poincare:1890}. However, little attention was paid to chaos until Edward Lorenz's work on nonlinear dynamics and description of butterfly effect in weather forecasting in the 1960s, which underpin the modern theory of deterministic chaos \cite{Lorenz:1963}.   Mathematically, a chaotic dynamics also exhibits dense periodic orbits and topologically mixing of its phase space open sets  \cite{Pecora:1997, GHu:1998}.  Chaos has been observed in a vast variety of realistic systems, including  Belouzov-Zhabotinski reactions,  nonlinear optics, Chua-Matsumoto circuit, Rayleigh-B\'{e}nard convention, meteorology, population dynamics,  psychology, economics, finance solar system, protein dynamics \cite{Braxenthaler:1997}, heart and brain of living organisms \cite{Boccaletti:2000}.  Various chaos control strategies have been proposed \cite{Ott:1990,Pecora:1990,Wei:2002e,Ashwin:2003}. However, the natural ability of  protein folding in controlling chaos  has not been unveiled yet.

Imagine that a folded protein is a  Greek chorus  where all particles sing with a synchronized voice on the dramatic action, while an unfolded protein is an anharmonic chaotic  orchestra where each particle plays its own rhythm with its own instrument. Mathematically, let us consider a folding protein that constitutes $N$ particles and has the spatiotemporal complexity of ${\mathbb{R}}^{3N}\times \mathbb{R}^{+} $. Assume that the molecular mechanics of the protein is described by a set of $N$ nonlinear oscillators of dimension ${\mathbb{R}}^{nN}\times \mathbb{R}^{+}$, where $n$ is the dimensionality of a single nonlinear oscillator. Although it is possible to consider all the physical interactions among protein particles, we feature the importance of  the protein distance geometry in this work. As such, we map a protein geometry into a set of topological relations or connectivities. The resulting interaction matrix defines the driving and response relation of  chaotic oscillators (see Supplementary Fig. \ref{fig:S1}). Amazingly, an $N$-time reduction in the spatiotemporal complexity  can be achieved, leading to an intrinsically low dimensional manifold (ILDM) of dimension ${\mathbb{R}}^{n}\times \mathbb{R}^{+}$. Without  loss of generality, we use 3D Lorenz attractors \cite{Lorenz:1963} to represent protein particles and their interaction matrix is built on molecular structures.

\begin{figure}
\begin{center}
\begin{tabular}{cc }
\includegraphics[width=0.44\textwidth]{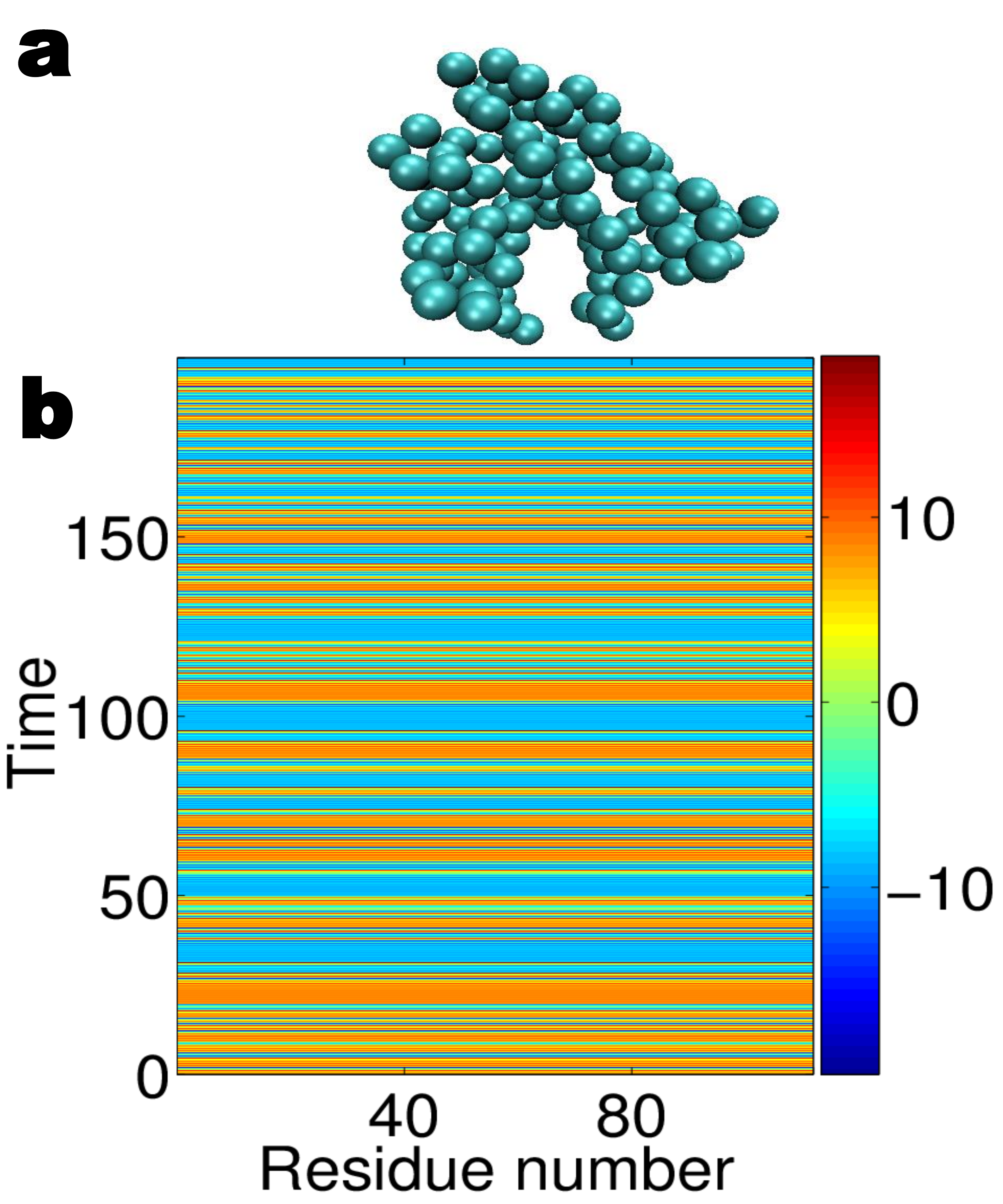}  &
\includegraphics[width=0.44\textwidth]{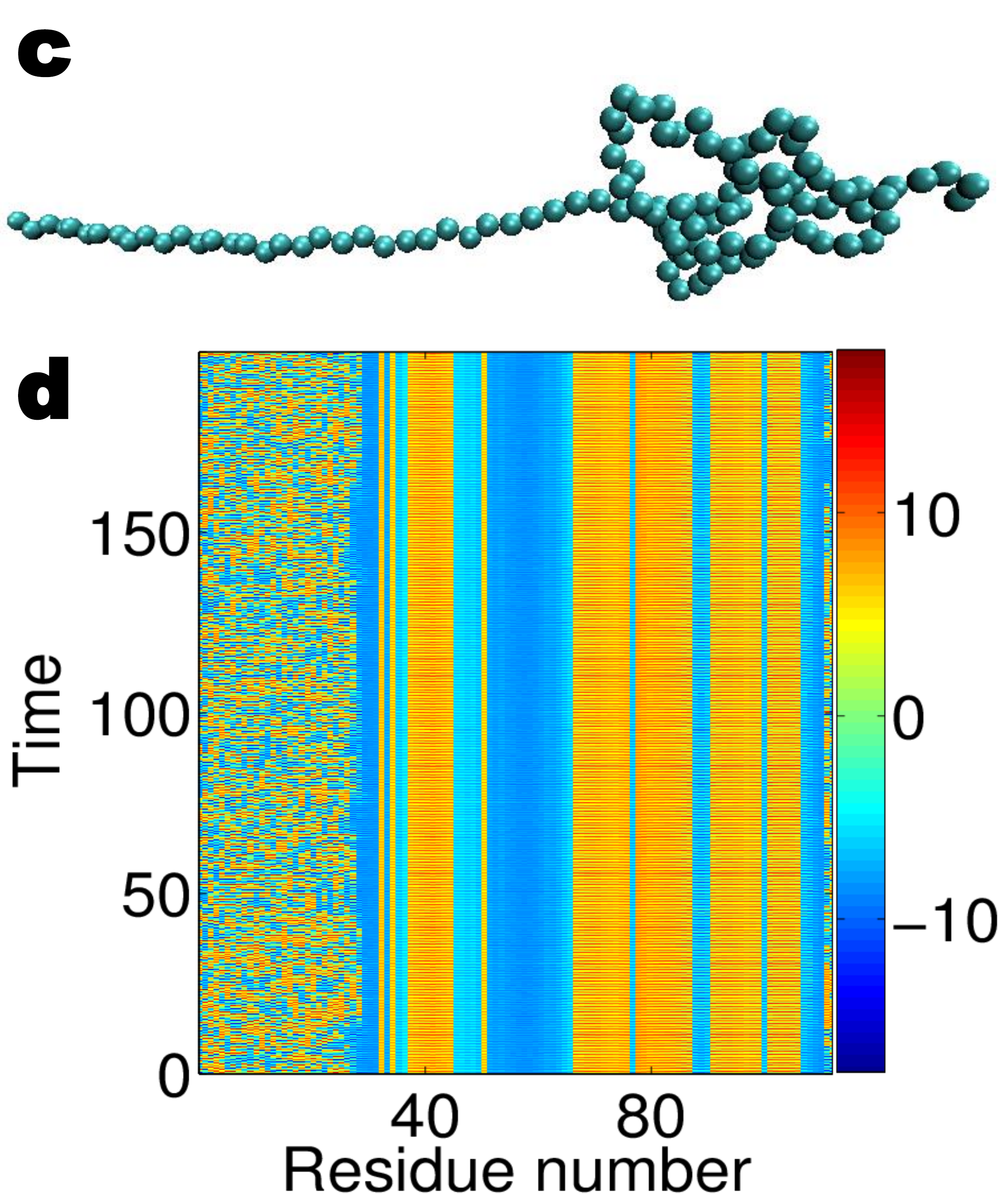} \\
\includegraphics[width=0.44\textwidth]{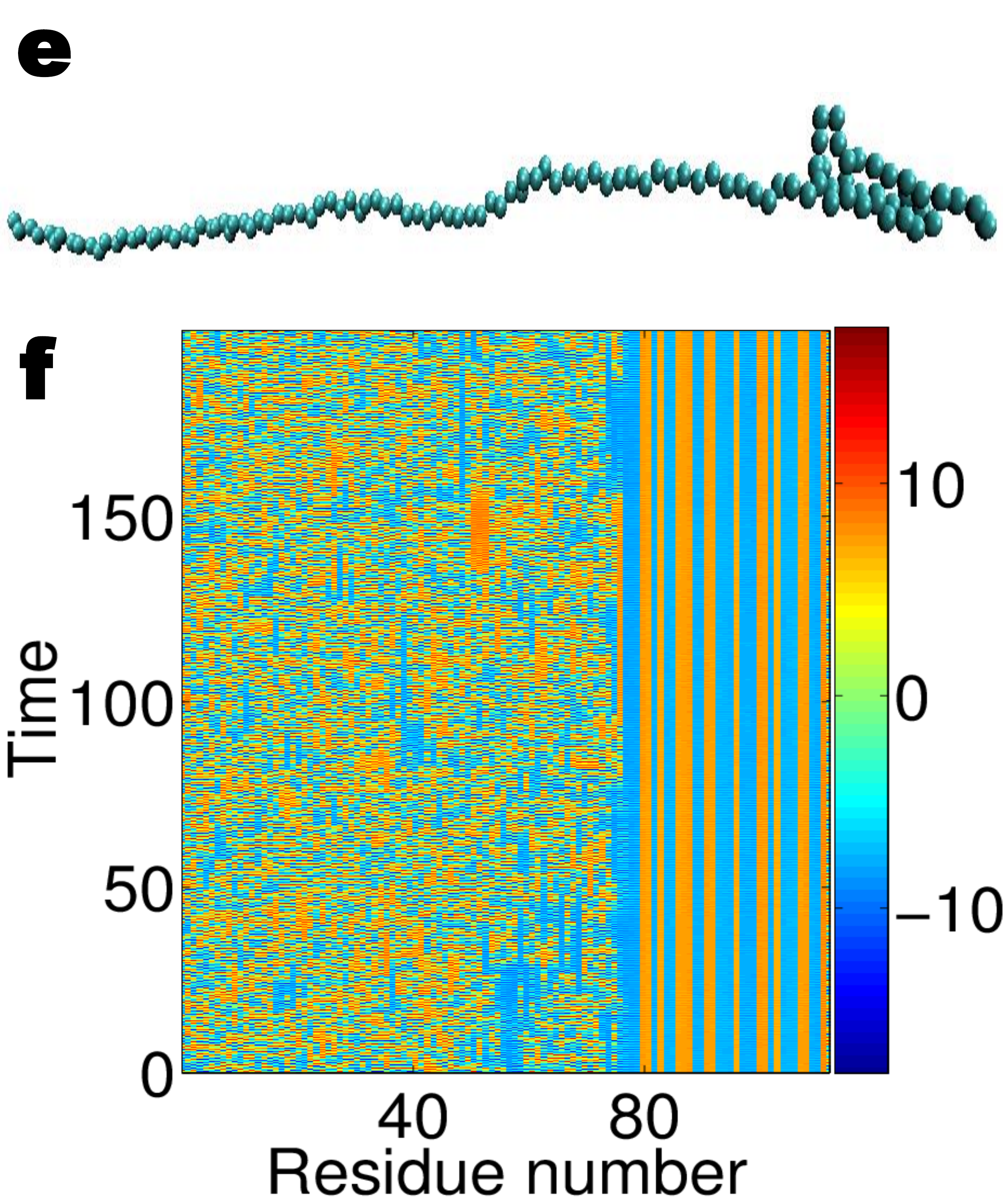} &
\includegraphics[width=0.44\textwidth]{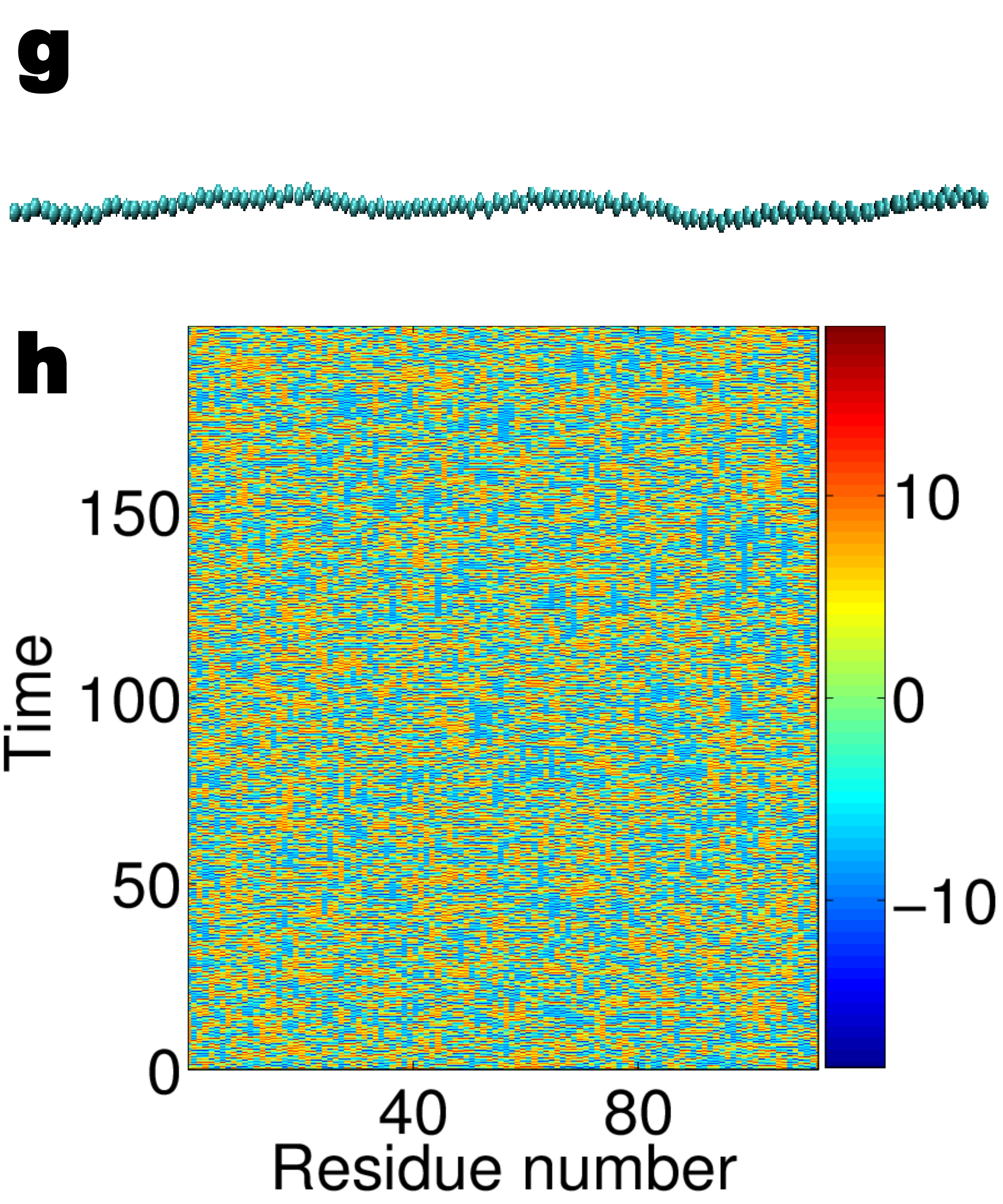}
\end{tabular}
\end{center}
\caption{The spectacular correlation between folded residues and controlled  chaos in the dynamics of macromomycin.
Forty partially folded or totally unfolded conformations are generated by pulling the folded structure (PBD ID: 2mcm).
{\bf a} The C$_\alpha$ atoms of the native folded structure.
{\bf b} The synchronous dynamics of a 3D ILDM for the native folded structure.
{\bf c} The C$_\alpha$ atoms of partially ordered Conformation 14.
{\bf d} The partial ordered chaotic dynamics of Conformation 14.  The first 30 residues are unfolded which leads to the chaotic dynamics in their nonlinear oscillators.
The dynamics of the last two residues are also chaotic for the same reason. Interesting   synchronized domains occur in the middle for partially folded residues.
{\bf e} The C$_\alpha$ atoms of more disordered Conformation 28.
{\bf f} The more chaotic dynamics of Conformation 28  showing a higher degree of randomness. Unfolded residues are in their chaotic motions. Compared to the dynamical behavior of Conformation 14, there are fewer synchronized domains and an average synchronized domain involves fewer oscillators.
{\bf g} The C$_\alpha$ atoms of completely unfolded Conformation 41.
{\bf h }  The completely chaotic dynamics of  Conformation 41 showing 336-dimensional chaotic motions.
}
\label{fig2}
\end{figure}

As a proof of principle, we first demonstrate that folded proteins are able to control chaos. To this end, we consider the  molecular nonlinear dynamics (MND) generated by the coarse-grained representation of bacteriocin AS-48 (protein data bank (PDB) ID: 1e68) by using its 70 amino acid residues. As a comparison, we create a reference dynamical system with 70 weakly coupled Lorenz attractors.  Parameters for both systems are the same and described in the Supplementary Materials. The same set of random data is used as initial conditions for both systems. The dynamics of  seventy weakly coupled Lorenz attractors  is chaotic as shown in Figs. \ref{fig1}{\bf a} and \ref{fig1}{\bf b}. In fact, each chaotic attractor resembles  the well-known wings of butterfly as plotted in  Fig. \ref{fig1}{\bf a}.  Surprisingly, the MND of bacteriocin AS-48  exhibits a shocking  transition from chaos to periodicity as depicted in Figs. \ref{fig1}{\bf c} and \ref{fig1}{\bf d}. The chaotic dynamics of each oscillator undergoes a  Hopf bifurcation as illustrated in Fig. \ref{fig1}{\bf c}.  It is interesting to note that there is a constant delay in the dynamics of any two adjacent oscillators and there is a lag synchronization in the protein dynamics. As such, the protein dynamics can be described by  an  ${\mathbb{R}}^{3}\times \mathbb{R}^{+}$ dimensional ILDM, which achieves a stunning $70$-fold reduction in  complexity and  dimensionality.

Having demonstrated the ability of transforming high dimensional chaos to a three dimensional periodic orbit  by a folded protein, we further analyze the dynamics of a set of forty one conformations generated from pulling the structure of macromomycin	(PDB ID: 2mcm) with a constant force. Generating unfolded proteins with a pulling force has been reported in the literature both computationally  \cite{Paci:2000} and experimentally \cite{Dudko:2006}. Conformation 1, shown in Fig. \ref{fig2}{\bf a}, is the folded    structure obtained by a short period of relaxation of the crystal structure from the PBD, while Conformations  2 to 41 are increasingly less folded due to the increase in the pulling force during their generation, see three typical ones in Figs.  \ref{fig2}{\bf c}, \ref{fig2}{\bf e} and \ref{fig2}{\bf g}.  The same set of random initial data is assigned to all the nonlinear oscillators in all conformations. Figure   \ref{fig2}{\bf b} shows the occurrence of a stable ILDM,  a synchronous chaos, in the dynamics of the folded structure. What spotlights the uniqueness and importance of the native structure is that none of any other conformations that are partially ordered and/or essentially disordered is able to tame chaos in their nonlinear dynamics under  the same condition. Indeed, the  dynamical systems of Conformations 2 to 41 are unstable,  chaotic,  and  $336$-dimensionally complex due to 112 residues. Consequently, they are very sensitive to initial values, simulation algorithms and time increments. However, it is still possible to extract some useful physical information from their unstable chaotic dynamics with well designed numerical experiments. Figures \ref{fig2}{\bf d} and   \ref{fig2}{\bf f}  indicate  synchronized domains in overall non-synchronized chaotic dynamics.  Amazingly, there is a miraculous correspondence between synchronized domains and the locations of partially folded amino acid residues,  which clearly indicates that protein folding leads to the control of chaos. It is interesting to note that synchronized domains are strikingly persistent over time. They  appear to be locked in certain ranges of solution values and show much less fluctuation and  volatility  than fully chaotic oscillators do.  Finally, as shown in Fig. \ref{fig2}{\bf h}, the dynamics of the completely unfolded  conformation is fully chaotic, which reinforces our observation that protein folding tames chaos and induces ILDM.

\begin{figure}
\begin{center}
\begin{tabular}{ccc}
\includegraphics[width=0.32\textwidth]{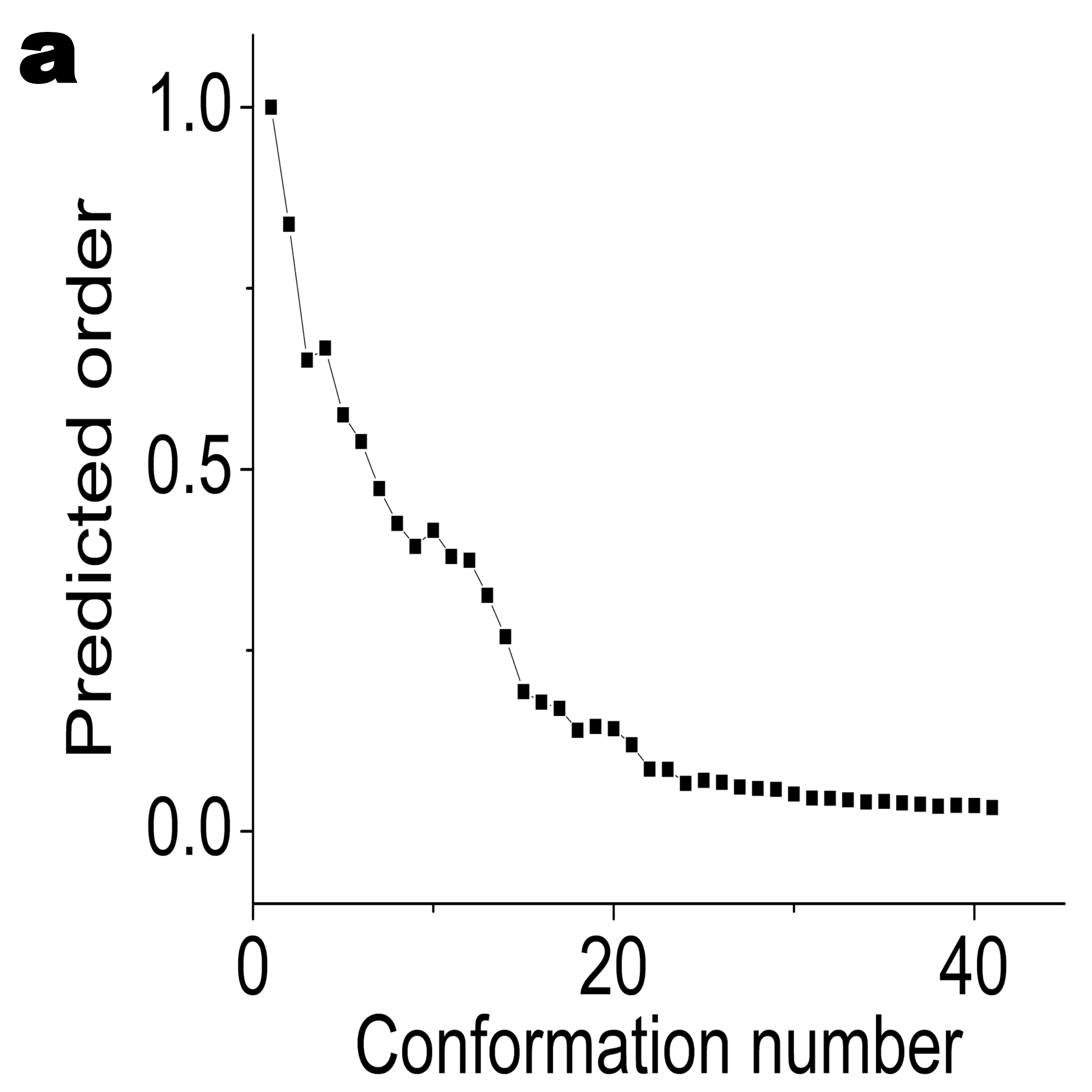} &
\includegraphics[width=0.32\textwidth]{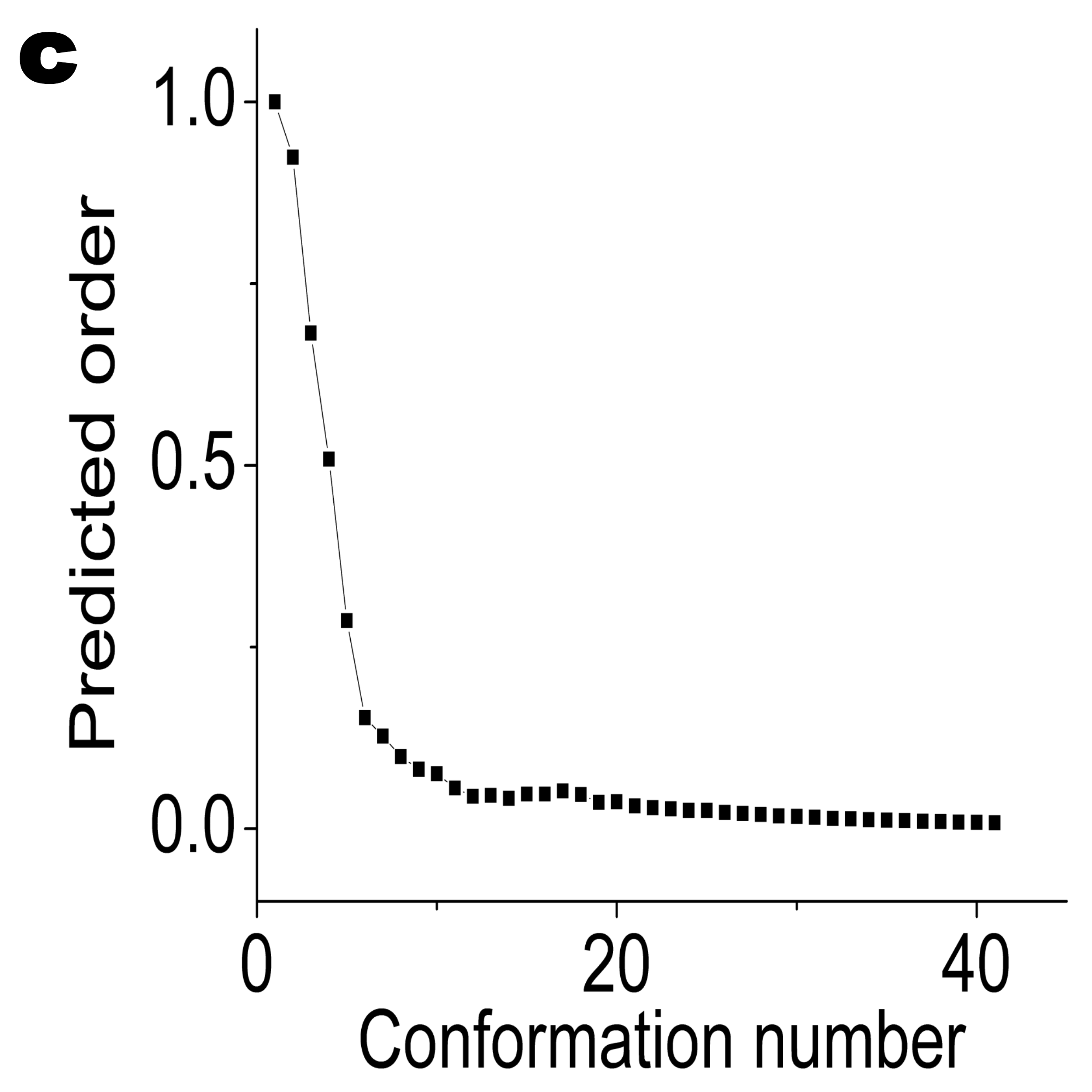} &
\includegraphics[width=0.32\textwidth]{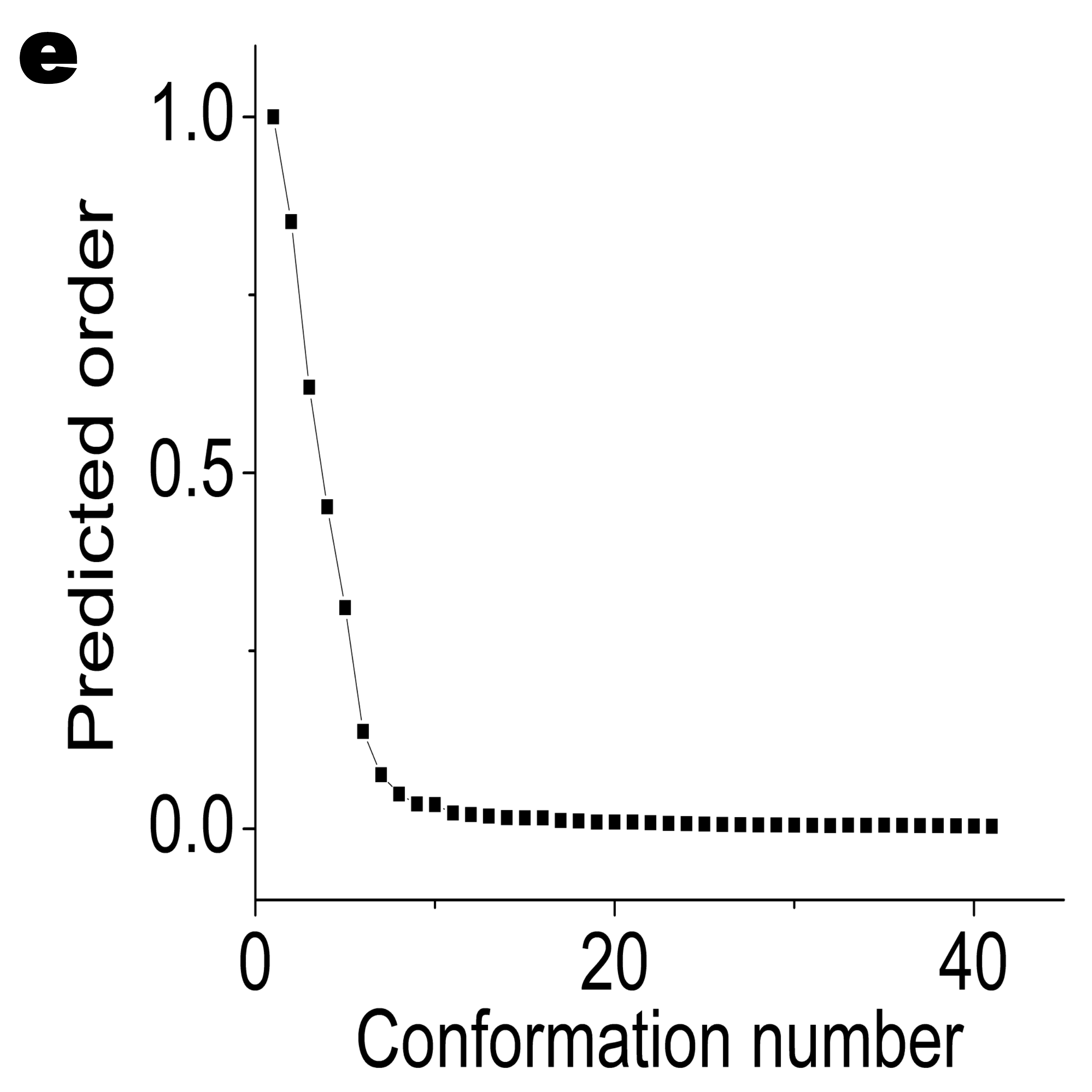} \\
\includegraphics[width=0.32\textwidth]{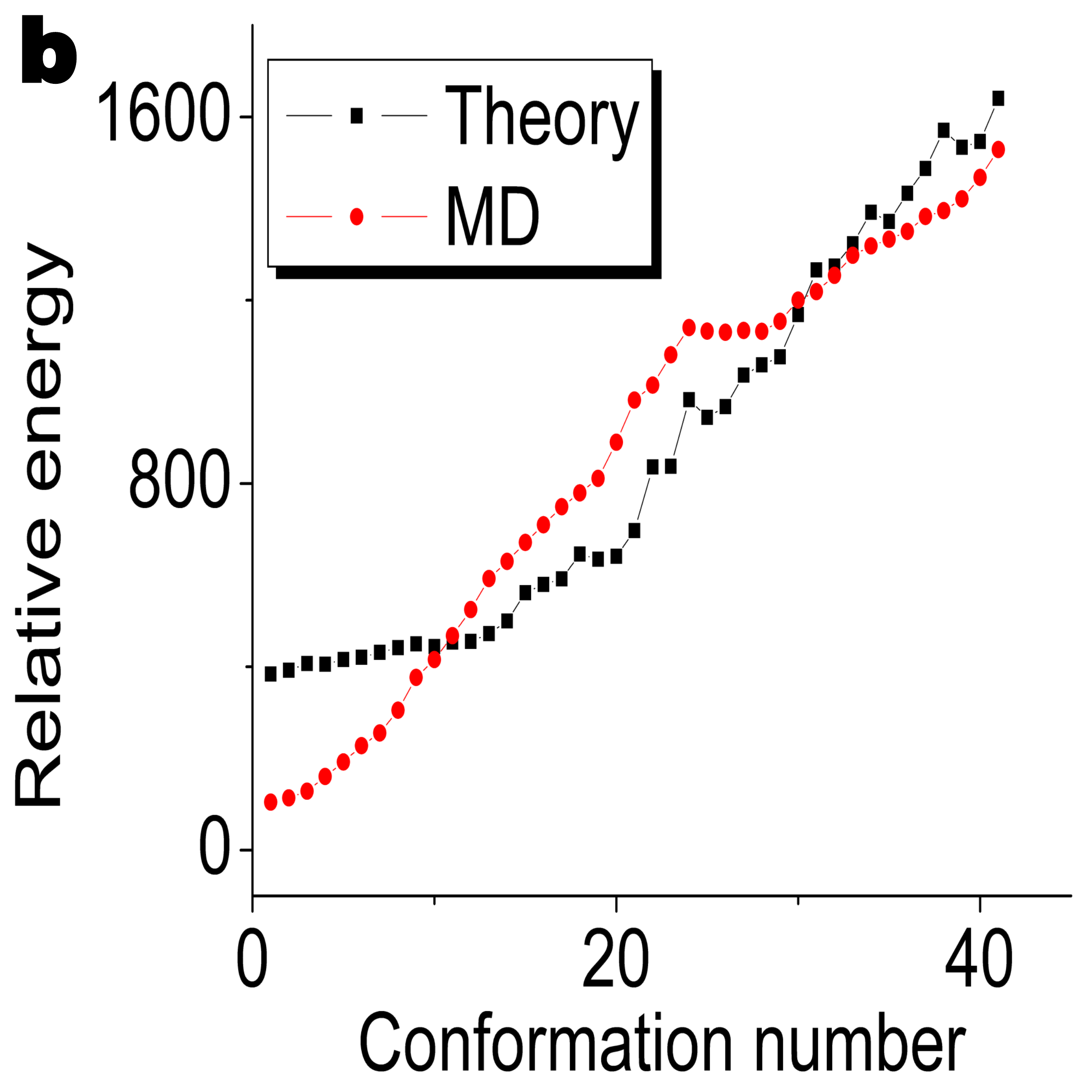} &
\includegraphics[width=0.32\textwidth]{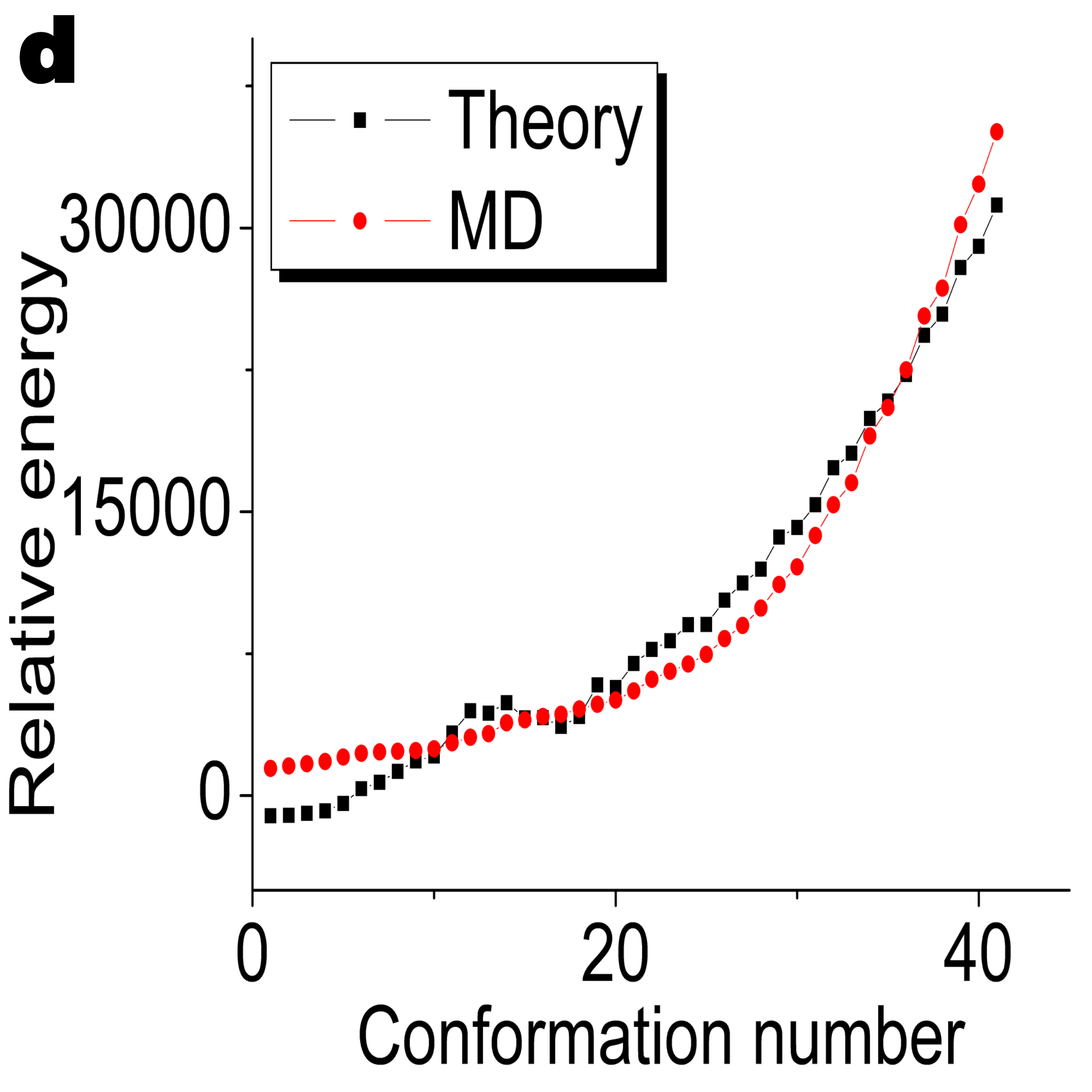} &
\includegraphics[width=0.32\textwidth]{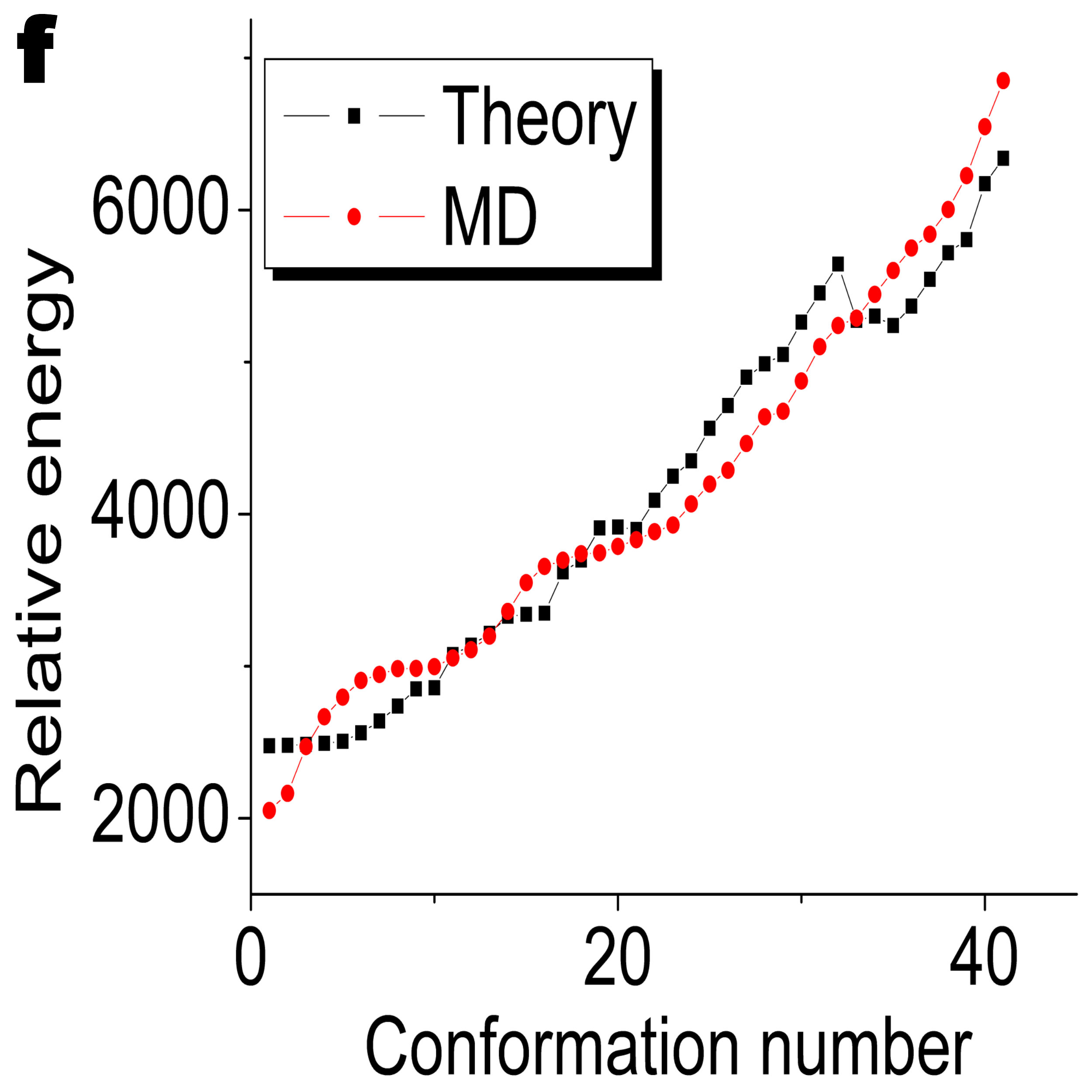}
\end{tabular}
\end{center}
\caption{ The orders and energy comparisons of three sets of protein conformations predicted by the stability analysis of the ILDM.
All conformations are generated from pulling native protein structures.
{\bf a} The predicted orders of  41 conformations for ubiquitin.
{\bf b} Comparison of relative energies of 41 conformations for ubiquitin.
{\bf c} The predicted orders of 41 conformations for phosphate-free bovine ribonuclease A.
{\bf d} Comparison of relative energies of 41 conformations for phosphate-free bovine ribonuclease A.
{\bf e} The predicted orders of 41 conformations for macromomycin.
{\bf f} Comparison of relative energies of 41 conformations for macromomycin.
 }
\label{fig3}
\end{figure}

To shed light on the mechanism of protein folding induced chaos control and ILDM, we analyze the transverse stability of the synchronous state. It turns out that the  stability problem of the  $n\times N$-dimensional nonlinear dynamics system is determined by its maximal Lyapunov exponent (MLE). Consequently, the ILDM is invariant with respect to a transverse perturbation if the MLE is smaller than zero. The MLE of a protein dynamics consists of two independent parts, i.e., the contribution from the single nonlinear attractor and that from the interaction matrix obtained from the protein distance geometry or the negative gradient of the protein interaction potential in general. The MLE of the single nonlinear attractor can be easily analyzed and is all known for the Lorenz attractor used in this work. While the contribution from protein distance geometry depends on the product of the interaction strength and the largest nonzero eigenvalue of the protein interaction matrix. The latter can be easily computed by a matrix diagonalization. As a result, there is a critical interaction strength for the chaotic dynamics of each protein conformation to arrive at its stable and invariant ILDM.

Interestingly, the above ILDM analysis gives rise to a new chaotic dynamics model (CDM) for the characterization of the disordered aggregation and the quantification of disorderliness in protein confirmations. A more disordered protein conformation requires a larger critical interaction strength to establish the synchronous state; whereas, the uniquely folded protein is, in principle, associated with the smallest critical interaction strength. To quantify  orderliness and disorderliness in protein conformations, we define an order parameter $\epsilon^n_c/\epsilon_c$, where $\epsilon_c$ and $\epsilon_c^n$ are the critical interaction strengths of a given conformation and the native conformation, respectively. Therefore, the order of the native protein conformation is 1 and that of a disordered protein is smaller than 1.  Figures \ref{fig3}{\bf a}, \ref{fig3}{\bf c}, and \ref{fig3}{\bf e}  illustrate the order parameters of  three sets of conformations generated from ubiquitin (PDB ID: 1ubq),  phosphate-free bovine ribonuclease A (PDB ID: 7rsa) and macromomyci (PBD ID: 2mcm). Forty partially folded or unfolded conformations are generated for each protein. Their order parameters exhibit a fast decay as their structures become less folded.

\begin{figure}
\begin{center}
\includegraphics[width=0.9\textwidth]{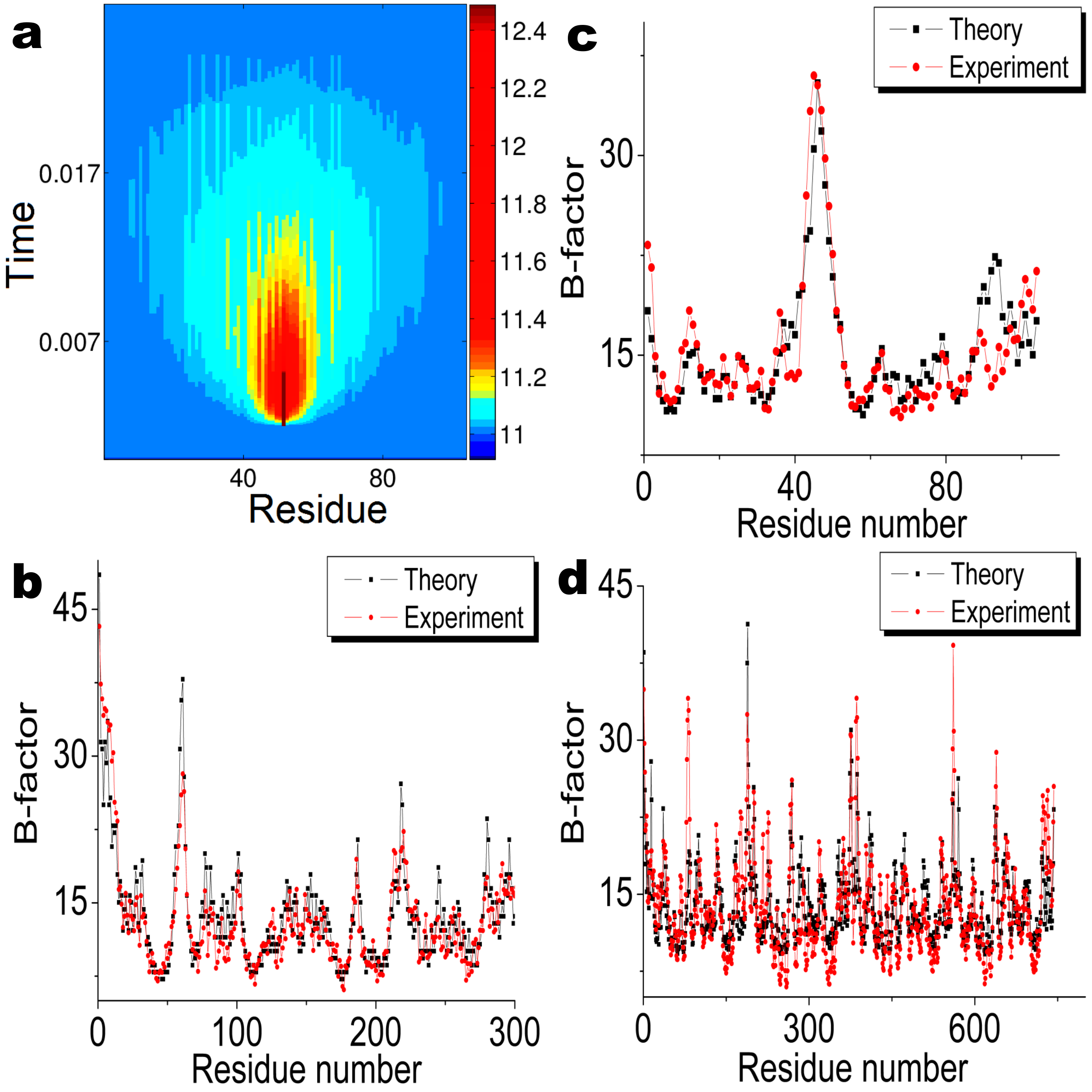}
\end{center}
\caption{Protein flexibility analysis by the present ILDM method.
{\bf a} The impact and relaxation of the transverse perturbation of the ILDM at a given amino acid residue (the 52th C$_\alpha$). The residues in horizontal axis are listed in the descending order according to their distances with respect to the perturbed residue.
{\bf b}-{\bf d} The experimental B-factors and ILDM predictions for protein  1aru, 2nuh and 4dr8. The correlation coefficients are respectively 0.913, 0.866  and 0.751 for three predictions.
}
\label{fig4}
\end{figure}

It remains to understand why the critical interaction strength is able to determine the  transverse stability of the chaotic dynamics of a given molecular conformation. As discussed in the  Supplementary Materials, the critical interaction strength is an inverse function of the second largest nonzero eigenvalue of the molecular interaction matrix. The latter is a manifestation of the molecular structure. As shown in Supplementary Fig. \ref{fig:S1}, the structure of a folded protein gives rise to a ``small-world'' interaction network \cite{Watts:1998}, while the structure of completely unfolded conformation has little nonlocal interactions. This implies that the  transverse stability of the ILDM  of a molecule is ultimately determined by its structure. Since the thermodynamical stability of a molecular structure is characterized by its total energy, there must be a one to one correspondence between the critical interaction strength and the total energy.  As such, we can estimate the relative energies of protein folding conformations based on their second largest nonzero eigenvalue of their interaction matrix. To verify this hypothesis, we analyze the nonlinear dynamics of protein structures 1ubq, 7rsa and 2mcm. For each structure,  we create a set of unfolded confirmations by molecular dynamics. Figures \ref{fig3}{\bf b}, \ref{fig3}{\bf d}, and \ref{fig3}{\bf f} show  good agreements between energies estimated by using the largest eigenvalues and those obtained from molecular dynamics simulations.

It is interesting to note that the intriguing dynamics of a protein ILDM is {\it exponentially} stable and persistently invariant. It will be wonderful if one can  take advantage of such properties in practical biophysical studies. To this end, we propose an ILDM based new method for protein rigidity analysis. Our essential idea is to perturb  the dynamics of each particle in a macromolecule in the transverse direction. Because of the stability of the ILDM, the nonlinear system must return to its original orbit, just like the free induction decay of the spin dynamics in NMR experiments. Similar to the T$_2$ and T$_1$ relaxations in an NMR experiment, the total relaxation time after the transverse perturbation, defined as the time used to recover the original state within a factor of $1/e$, is a measure of the strength of its particle-particle   and particle-environment interactions. For a given particle,   stronger interactions with neighboring particles and environment lead to a shorter relaxation time, which translates into higher rigidity and lower Debye-Waller factor, or   B-factor. Therefore, the direct connection between the thermodynamical stability and the ILDM stability enables us to  quantitatively estimate atomic temperature factors in a molecule.

Figure \ref{fig4}{\bf a} illustrates the relaxation process of a perturbed nonlinear dynamics of protein 2nuh. The instantaneous perturbation is propagated from the nearest neighboring amino acid residues to a wider region over a time period before gradually fades off. By recording the relaxation time, one is able to predict the B-factor of an amino acid residue and compare it to the experimental data given by X-ray crystallography.  Figures  \ref{fig4}{\bf b}- \ref{fig4}{\bf d} provide such  comparisons for three protein structures, namely 2nuh, 1aru and 4dr8.  It is seen that our results obtained from the perturbation of the ILDM  are in a very good consistency with those of X-ray data. To   demonstrate the robustness of the present chaotic dynamics model, we have listed the correlation coefficients of our predictions and X-ray data for a set of 30 proteins in Supplementary Table \ref{table:S1}. A comparison with the state of the art GNM for B-factor prediction depicted in Supplementary Fig. \ref{fig:S5} further validates the proposed ILDM method.

It is noted that the proposed chaotic dynamics model  can also be used to quantify the relative rigidity of protein folding conformations generated by a constant pulling force. As shown in Supplementary Fig. \ref{fig:S6}{\bf a},  the relative B-factors of six conformations of macromomycin show the flexibility at both the C terminal and the N terminal. The middle section is relatively rigid against a constant pulling force.  Residues 90-99 appear to be the most rigid ones. The persistence of folded residues can be analyzed by averaged relative  B-factors obtained from all of forty one conformations as shown in  Supplementary Fig. \ref{fig:S6}{\bf b}. Although normal B-factors indicate the flexibility of each residue, the averaged relative  B-factors illustrate the dynamics stability of each residue under external pulling forces, which might be used to analyze interaction potentials for individual residues.

Finally, our ILDM  based prediction of B-factors can be improved by a more detailed description of neighboring amino acid residues and other adjacent cofactors, such as metal clusters, binding ligands, and proteins. Although a throughout investigation is out of the scope of the present work, we illustrate the improvement of our prediction by the inclusion of two Fe$_4$S$_4$ metal clusters in  ferredoxin (PDB ID: 1fca). The chaotic dynamics of  two Fe$_4$S$_4$ atomic oscillators serves as the driving source and the chaotic  dynamics of all neighboring amino acid residues is considered as the response system. Indeed, the simple consideration of 16 Fe$_4$S$_4$ metal clusters gives rise to near five percent improvement in the correlation coefficients as shown in  Supplementary Fig. \ref{fig:S7}.

\section{ Supplementary information}

\subsection{Geometry to topology mapping} \label{Sec:GTM}

Topological relations or connectivities among molecular particles are needed in the molecular chaotic dynamics. This topological information can be extracted from the geometric properties of a given molecule of interest. Let us consider a molecule of $N$ particles located at ${\bf r}_1, {\bf r}_2,\cdots, {\bf r}_{N}$, where ${\bf r}_j \in {\mathbb{R}}^3$, where particles are either atoms, amino acids residues or other superatoms in the molecule. The distance between the $i$th and $j$th particles is given by $d_{ij}({\bf r}_i,{\bf r}_j)=\|{\bf r}_i-{\bf r}_j\|_2$. The interaction matrix must satisfy the driven and response relation between two dynamics systems \cite{Pecora:1997,GHu:1998}. Additionally, we assume that all particles are mutually connected and their interactions decay as a function of their distance ${\bf A}_{ij}(d_{ij})$.  The simplest form for the interaction matrix is the Kirchhoff (or connectivity) matrix generated by cutoff distances $\sigma_{ij}$
\begin{eqnarray}\label{eq:couple_matrix3}
{\bf A}_{ij} &=& \left\{
\begin{array}{ll}
1, & \forall  d_{ij}\leq \sigma_{ij},  i\neq j\\
0, &  \forall d_{ij} >   \sigma_{ij},  i\neq j\\
 -\sum_{j\neq i}{\bf A}_{ij}, & \forall  i=j.
\end{array}
\right.
\end{eqnarray}
To account for the distance effect in a more realistic manner, it is also convenient to employ  monotonically decreasing radial basis functions or delta sequence kernels of positive type \cite{GWei:2000}. Here we consider   generalized exponential functions
\begin{eqnarray}\label{eq:couple_matrix0}
{\bf A}_{ij} = \left\{\begin{array}{ll}
 e^{- d^k_{ij}/k\sigma^k_{ij}}, & \quad \forall i\neq j,  \quad k=1,2,\cdots \\
 -\sum_{j\neq i}{\bf A}_{ij}, & \quad \forall i = j
\end{array}
\right.,
\end{eqnarray}
and   power-law functions
\begin{eqnarray}\label{eq:couple_matrix1}
{\bf A}_{ij} = \left\{\begin{array}{ll}
 (d_{ij}/\sigma_{ij})^{-\upsilon} , & \quad \forall i\neq j,  \quad \upsilon>1,\\
 -\sum_{j\neq i}{\bf A}_{ij}, & \quad \forall i = j.
\end{array}
\right.,
\end{eqnarray}
where $\sigma_{ij}$ are   characteristic distances between particles. In the present model, $\sigma_{ij}$ can be used as a  set of fitting parameters. Note that some of the above matrix expressions have  also been used in other flexibility analysis approaches \cite{Bahar:1997,Bahar:1998,Atilgan:2001,Hinsen:1998,Tama:2001,LiGH:2002}. However, the present construction of these functional forms was based on the driven and response relation of coupled dynamical systems \cite{Pecora:1997}.

\begin{figure}
\begin{center}
\begin{tabular}{cc}
\includegraphics[width=0.5\textwidth]{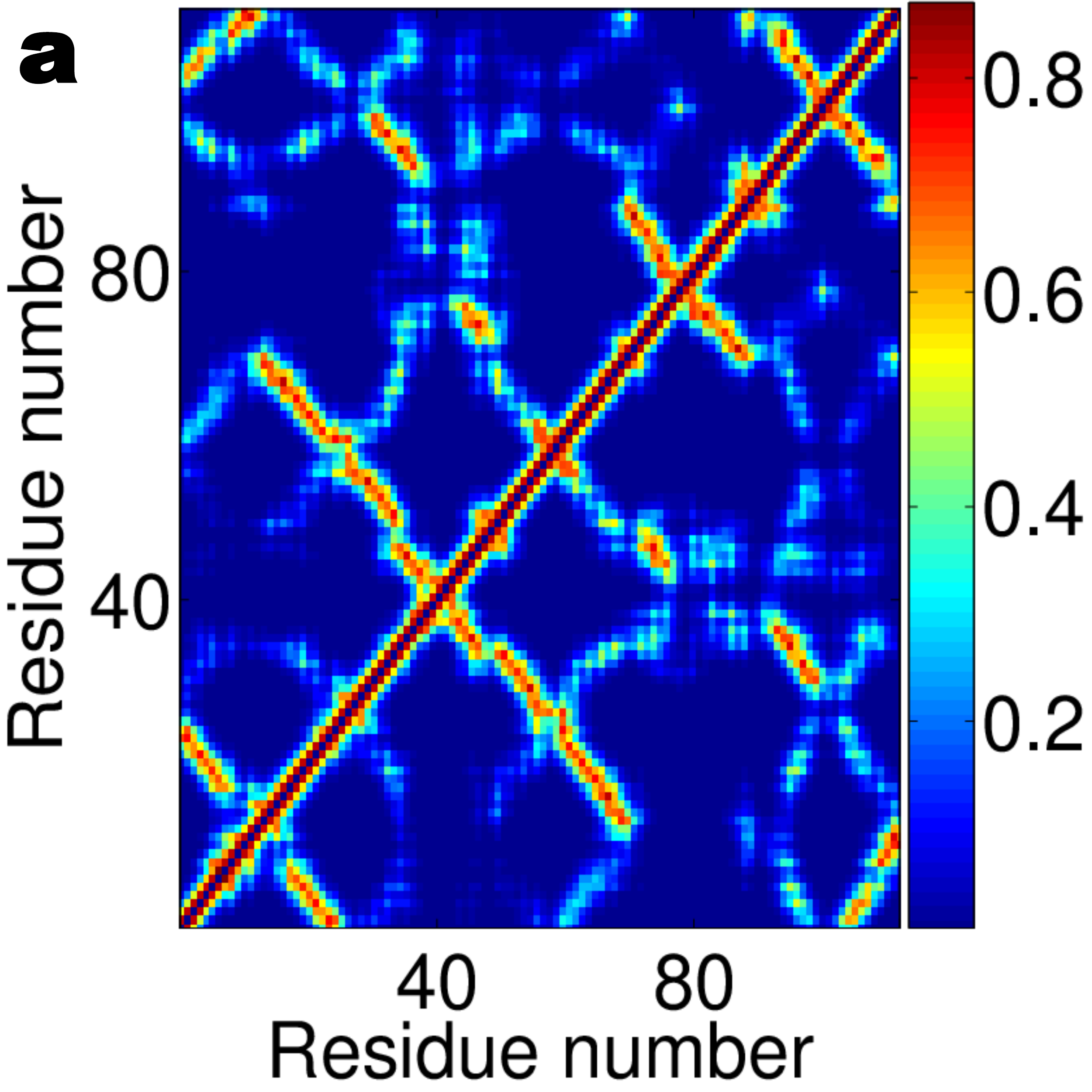}
\includegraphics[width=0.5\textwidth]{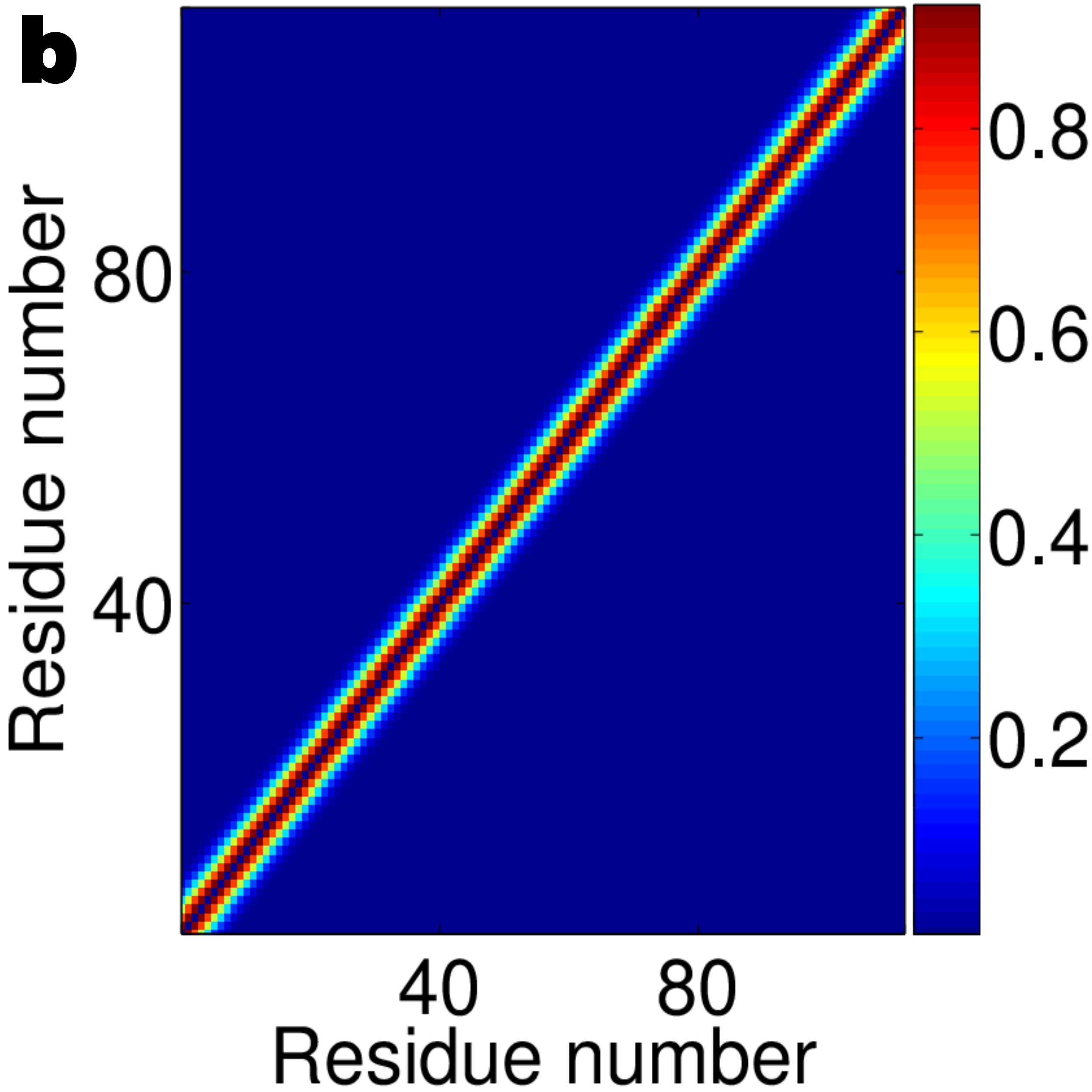}
\end{tabular}
\end{center}
\caption{  The  interaction matrices 2 conformations of protein 2mcm  generated by using  Eq. (\ref{eq:couple_matrix0}) with $k=2$ and $\sigma_{ij} =6$. Diagonal elements have been excluded to emphasize non diagonal interactions.
{\bf a} The interaction matrix of the native structure of 2mcm shown in  Fig. 2{\bf a} indicates much nonlocal interactions.
{\bf b} The interaction matrix of a completely unfolded conformation shown in Fig. 2{\bf g} demonstrates little nonlocal interactions.
}
\label{fig:S1}
\end{figure}

Expressions (\ref{eq:couple_matrix3})-(\ref{eq:couple_matrix1})  map a molecular geometry into topological relations or connectivities. The interaction matrix ${\bf A}$ is an $N\times N$ symmetric, diagonally dominant matrix.  Note that elements in the interaction matrix are not interaction potentials among particles. Except  specified, Eq. (\ref{eq:couple_matrix0}) with $k=2$ is used in the present work. For simplicity, we set characteristic distances to $\sigma_{ij}=\sigma$ for all amino acid residues.    Figure \ref{fig:S1} illustrates the non-diagonal elements of two interaction matrices, one for the native protein structure of 2mcm and the other for completely unfolded conformation generate from pulling the structure of 2mcm. It is seen that the native  protein shows a heterogeneity (or small-world property \cite{Watts:1998}) in its interaction matrix, which contributes to the protein stability. Whereas the interaction matrix of the completely unfolded  conformation has little nonlocal interaction.

\subsection{Stability analysis of the ILDM}\label{Sec:Stability}

Understanding deterministic chaos is of theoretical and practical importance \cite{Ott:1990,Pecora:1990,Wei:2002e,Ashwin:2003}.
Some detailed analysis of chaos dynamics can be found in the literature \cite{Pecora:1997,GHu:1998}. Here, we provide a brief discussion of the theoretical foundation used in the present work.
Let us consider an $n\times N$-dimensional nonlinear system for $N$ interacting chaotic oscillators
\begin{eqnarray}
\label{eq:couple_matrix}
\frac{d{\bf u}}{dt} &=& {\bf F}({\bf u})+ {\bf E}{\bf u}, ~~~
\end{eqnarray}
where ${\bf u}=({\bf u}_1,{\bf u}_2,\cdots, {\bf u}_N )^T$ is an array of state functions for $N$ nonlinear oscillators, ${\bf u}_j=(u_{j1},u_{j2}, \cdots, u_{jn})^T$ is an $n$-dimensional nonlinear function for the $j$th oscillator,   $ {\bf F}({\bf u})=(f({\bf u}_1),\\
f({\bf u}_2), \cdots, f({\bf u}_N))^T$ is an array of nonlinear functions of $N$  oscillators, and ${\bf E}=\varepsilon{\bf A}\otimes \Gamma$.  Here,  $\varepsilon$ is the overall  interaction strength,  ${\bf A}$ is the $N\times N$ interaction matrix defined in Section \ref{Sec:GTM} and $\Gamma$ is an $n\times n$ linking matrix.  The $n$-dimensional ILDM is defined as
\begin{eqnarray}
{\bf u}_1(t)={\bf u}_2(t)=\cdots = {\bf u}_N(t)={\bf s} (t),
\end{eqnarray}
where ${{\bf s} (t)}$ is a synchronous state or reference state.

To understand the stability of the ILDM of protein chaotic dynamics, we define a transverse state function as ${\bf w}(t)={\bf u}(t)-{\bf S}(t)$, where ${\bf S}(t)$ is a vector of $N$ identical components $({\bf s}(t),{\bf s}(t),\cdots, {\bf s}(t) )^T$. Obviously, the invariant  ILDM is given by  ${\bf w}(t)={\bf u}(t)-{\bf S}(t)={\bf 0}$. Therefore, the stability of the ILDM can be analyzed by $\frac{d {\bf w}(t)}{dt} =\frac {d {\bf u}(t)}{dt} -\frac {d {\bf S}(t)}{dt}$, which can be studied by the following linearized equation
\begin{eqnarray}\label{eqn:trans}
\frac{d{\bf w}}{dt} = ({\bf DF} ({\bf s}) + {\bf E}){\bf w}, ~~~
\end{eqnarray}
where ${\bf DF}({\bf s})$ is the Jacobian of ${\bf F}$.

To further analyze the stability of Eq. (\ref{eqn:trans}), we introduce  a set of $N$ vectors $\{\phi_j\}_{j=1}^N$ that diagonalize the interaction matrix ${\bf A}$
\begin{eqnarray}\label{eqn:trans2}
{\bf A}\phi_j(t)  =\lambda_j \phi_j(t), \quad  j=1,2,\cdots, N.
\end{eqnarray}
The transverse state vector has the expansion of
\begin{eqnarray}
{\bf w}(t)=\sum_{j}{\bf v}_j(t)\phi_j(t).
\end{eqnarray}
The stability problem of the ILDM  is equivalent to the following  stability problem
\begin{eqnarray}\label{eqn:trans3}
\frac{d{\bf v}_j(t)}{dt} &=& ( Df ({\bf s}) + \varepsilon \lambda_j\Gamma){\bf v}_j(t), \quad j=1,2,\cdots,N,
\end{eqnarray}
where $Df ({\bf s})$ is the diagonal component of ${\bf DF} ({\bf s})$.
The stability of Eq. (\ref{eqn:trans3}) is determined by the largest Lyapunov exponent $L_{\rm max}$, namely, $L_{\rm max} < 0$, which can be decomposed into two contributions
$$L_{\rm max}=L_{\rm f}+L_{\rm c},$$
where $L_{\rm f}$ is the largest  Lyapunov exponent of the original $n$ dimensional  chaotic system $\frac{d{\bf s}}{dt} =  { f}({\bf s})$, which can be easily computed for most chaotic systems. Here, $L_{\rm c}$ depends on   $\lambda_j$ and $\Gamma$. The largest eigenvalue  $\lambda_1$ equals  0,  and its corresponding eigenvector represents the homogeneous motion of the ILDM, and all of other eigenvalues $\lambda_j, j=2,3,\cdots, N$ govern the transverse stability of the ILDM. Let us consider a simple case in which the linking matrix is the unit matrix ($\Gamma={\bf I}$). Then stability of the ILDM is determined by  the second largest eigenvalue $\lambda_{2}$, which enables us to estimate the critical interaction strength $\varepsilon_{c}$   in terms of $\lambda_{2}$ and $L_{f}$,
\begin{equation} \label{eq:expect}
\varepsilon_{c}=\frac{L_f}{- \lambda_{2}}.
\end{equation}
The dynamical system reaches the ILDM when $\varepsilon > \varepsilon_{c}$ and is unstable when $\varepsilon \leq \varepsilon_{c}$.
The eigenvalues of protein interaction matrices  are obtained with a standard matrix diagonalization algorithm.

\subsection{Lorenz attractor and its parametrization}

\subsubsection{Lorenz equations}
For simplicity, we choose a set of  $N$ Lorenz attractors \cite{Lorenz:1963} to illustrate our ideas. The Lorenz equation is  three-dimensional
 $\mathbf{u}_i=(x_i,y_i,z_i)^T$,
\begin{eqnarray}  \nonumber
   \frac{dx_i}{dt}&=&\alpha(y_i-x_i) \\  \label{oscillator}
   \frac{dy_i}{dt}&=&\gamma x_i-y_i-x_iz_i \\ \nonumber
   \frac{dz_i}{dt}&=&x_iy_i-\beta z_i, i= 1, 2, \cdots, N
\end{eqnarray}
where parameters $\alpha>0, \beta>0$ and $\gamma>0$ are to be specified for each given system.

The Lorenz equation (\ref{oscillator}) can be analyzed with the Poincar\'{e} section and the first return map \cite{Tucker:2002}. Parameter $\gamma $ has been used to classify certain behavior of the Lorenz equation (\ref{oscillator}). The origin is a fixed point when $\gamma < 1$.  When $\gamma = 1$, the system is at the saddle-node bifurcation point.   For $\gamma > 1$, there is a pair of  fixed points given by
$$
{\bf u}^{\pm}=(\pm \sqrt{\beta(\gamma-1)}, \pm \sqrt{\beta(\gamma-1)}, \gamma-1)^T.
$$

For $\alpha>\beta+1$, the above twin fixed points are stable if
$$
\frac{\gamma}{\alpha}<\frac{\alpha+\beta+3}{\alpha-(\beta+1)}.
$$
At  $\alpha = \beta+1$ the twin fixed points are no longer stable due to the Hopf bifurcation.

By choosing the classical parameter values $\alpha=10,\beta=8/3 $ and $\gamma=28$,  almost all points in the phase space go to a strange attractor - the Lorenz attractor.
The Lorenz equations are solved by using the forward Euler scheme and/or  the fourth order Runge-Kutta scheme.

\subsubsection{Parameters used in protein induced Hopf bifurcation  from chaos }

Figure 1 illustrates the transition from chaos to periodicity induced by protein 1e68. There are 70 amino acid residues in 1e68 and therefore, we have 70 chaotic oscillators ($N=70$). Figure 1{\bf a} is the original chaotic system with a special matrix without any realistic protein interaction. In this case,  we set ${\bf E}$  in Eq. (\ref{eq:couple_matrix}) as ${\bf E}=\eta{\bf B}\otimes \Gamma$, where ${\bf B}$ is given by
 \begin{eqnarray}\label{eq:couple_matrix4}
{\bf B}_{ij} &=& \left\{
\begin{array}{rl}
1, & \forall  d_{ij}\leq \sigma_{ij},  j>i\\
-1, & \forall  d_{ij}\leq \sigma_{ij}, j<i\\
0, &    {\rm otherwise}
\end{array}
\right..
\end{eqnarray}
The linking matrix in Fig. 1 is given as
\begin{eqnarray}
\Gamma= \left( \begin{array}{ccc}
                      0               & 0             & 0 \\
                      1               & 0             & 0 \\
                      0               & 0             & 0
                      \end{array}
                      \right).
\end{eqnarray}
The linking matrix is chosen as ${\bf I}$, the unit matrix, in all the other cases.

In Figs. 1{\bf c} and 1{\bf d}, additional protein interactions are considered.  In this case,  we set ${\bf E}=(\varepsilon{\bf A}+\eta{\bf B})\otimes \Gamma$,  where the interaction matrix ${\bf A}$ takes the form given in   Eq. (\ref{eq:couple_matrix3}).  Here, $\eta=7$, $\varepsilon=10$ and  $\sigma_{ij}=4$\AA~ are used.  The Lorenz parameters are set to   $\alpha=1$, $\gamma=60$ and $\beta=\frac{8}{3}$.



\subsection{Generation of unfolded conformations using steered molecular dynamics}
The folding and unfolding mechanism of  proteins is of fundamental significance to structure biology. Experimental tools such as atomic force microscopy and optical tweezers are often employed to explore the mechanism with a single molecule. Due to the limitations of  experimental means, steered molecular dynamics (SMD) has been developed and established as a computational tool for the study of protein structure and dynamics. With this approach, the folding/unfolding pathway of the ligand-receptor and elasticity behavior of the muscle proteins are extensively studied \cite{Paci:2000,Isralewitz:2001,Gao:2002,Hui:1998,Srivastava:2013}.

Computationally, an unfolding process can be realized through one of three ways: high temperature, constant force pulling , and constant velocity pulling \cite{Paci:2000,Hui:1998,Srivastava:2013}. Implicit and explicit solvent models have been used in SMD simulations \cite{Paci:1999,Paci:2000}. In the study of the mechanical property of protein FN-$III_{10}$, it is believed that implicit solvent models tend to miss the friction due to water molecules and the effect of explicit water molecules on breaking inter-strand hydrogen bonds \cite{Gao:2002}. To a certain extend, explicit solvent models with a large water sphere can be also incomplete, as the deformation of the water sphere requires additional artificial force. Even when the spherical periodical boundary condition is employed, the extension tends to go beyond the boundary of the sphere. It is argued that an explicit solvent model with a large box which solvates the stretched protein can be more reliable \cite{Gao:2002}. However, this simulation requires extensive computational resource.

In the present work, we explore the chaoticity and orderliness of the folded, partially folded and unfolded protein conformations. To simplify the relation between protein energy and conformation, and between protein orderliness and conformation, we  carry out SMD simulations   without water molecules. To demonstrate that our findings are valid in general, we perform  additional simulations with a large water sphere similar to the SMD setting in the literature \cite{Hui:1998}.

\begin{figure}
\begin{center}
\begin{tabular}{c}
\includegraphics[width=0.8\textwidth]{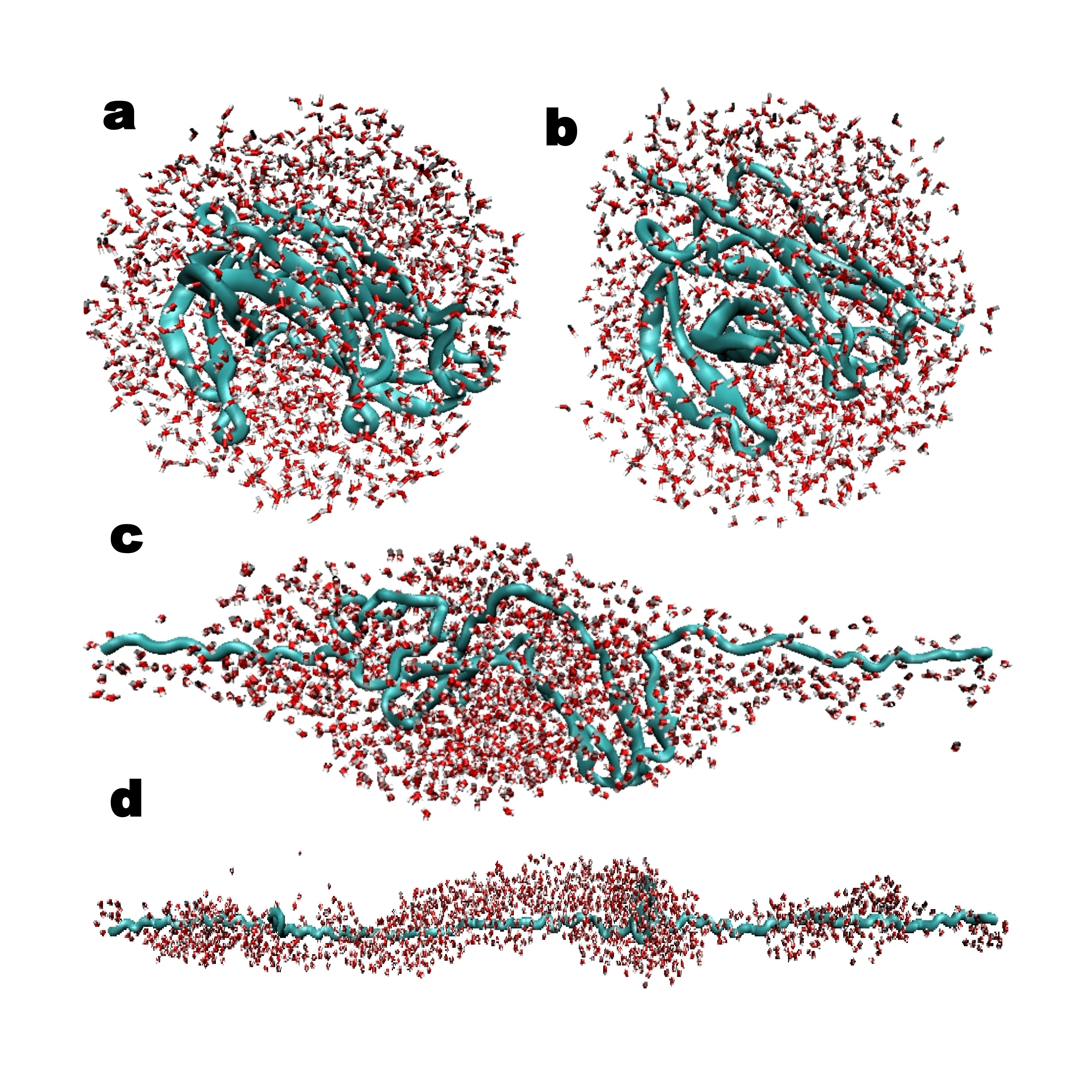}
\end{tabular}
\end{center}
\caption{ Four typical conformations of protein 2mcm generated with  SMD simulations with water molecules. {\bf a}
The initial native protein structure (Conformation 1) and its water environment. {\bf b, c} and {\bf d} Three conformations (2, 11, and 31) generated by using constant velocity pulling.
}
\label{fig:S2}
\end{figure}
\subsubsection{SMD simulations  without water molecules}

It is quite standard to create partially folded and unfolded proteins by a pulling force in computer simulations \cite{Paci:2000} and with experimental means \cite{Dudko:2006}. The relation between protein topology and energy was discussed in the literature \cite{Paci:2000}. In  Figs. 2, 3 and \ref{fig:S6}, partially folded and unfolded protein conformations are obtained by using the molecular dynamical simulation tool NAMD (http://www.ks.uiuc.edu/Training/Tutorials/namd/namd-tutorial-html/) with a  constant pulling velocity.  In the protein preparation procedure, a protein structure downloaded from the PDB is first submerged into a  box with a layer of 5{\AA}~ water in each direction from the atom with the largest coordinate in that direction. We use the time interval of 2ps in our simulations. A total of 15000 time step equilibration is performed with  the periodic boundary condition after 10000 time step initial energy minimization.

It is no doubt that water molecules are of significant importance for protein folding pathways and  biological functions. However, since the purpose of the present work is to reveal the relation between protein structures and their chaotic dynamics, water molecules and their impact to the protein conformations are somewhat irrelevant to our findings. Therefore, we construct partially folded and unfolded protein conformations by pulling the relaxed protein structures in SMD simulations with a constant velocity.  Basically, we fix the first C$_\alpha$ atom and apply a constant pulling velocity on the last C$_\alpha$ atom along the direction that connects these two atoms. We set the spring constant as 7 kcal/mol{\AA}$^2$, while 1 kcal/mol{\AA}$^2/$  equals   69.74 pN {\AA}. The constant velocity is 0.005{\AA}~ per time step. A total of 20000 simulation steps is integrated in generating 1ubq conformations and a new  conformation is extracted after every 500 simulation steps. For proteins 7rsa and 2mcm, a total of 80000 simulation steps is employed  for each  protein and new  conformations are extracted at the frequency of every 2000 steps.  To verify our findings, we have varied the procedure of conformation generation in many ways, including using different initial structure preparations,  number of integration steps, and pulling velocities, which lead to different sets of conformations. However, our findings presented in Figs. 2, 3 and \ref{fig:S6} are  not affected by these variations.

\paragraph{Parameters}
In Fig. 2, we set the characteristic distance $\sigma_{ij}=10$\AA~ and the interaction strength $\varepsilon=0.12$. For  parameters in the Lorenz dynamic system, we use $\alpha=10$, $\gamma=28$ and  $\beta=\frac{8}{3}$. Random numbers of   range [0,1] are used for as initial conditions for all oscillators. The forward Euler scheme with the time increment of $h=10^{-2}$ is used for the time integration.

In Figure 3,  $\sigma_{ij}=4$\AA,  $\alpha=10$, $\gamma=28$ and $\beta=\frac{8}{3}$ are used.  The maximal Lyapunov exponent for each oscillator is about 0.908. The critical interaction strength is  $\varepsilon_{c}=\frac{0.908}{- \lambda_{2}}$. Here $\lambda_{2}$ can be evaluated once the interaction matrix is constructed.

\begin{figure}
\begin{center}
\begin{tabular}{cc}
\includegraphics[width=0.45\textwidth]{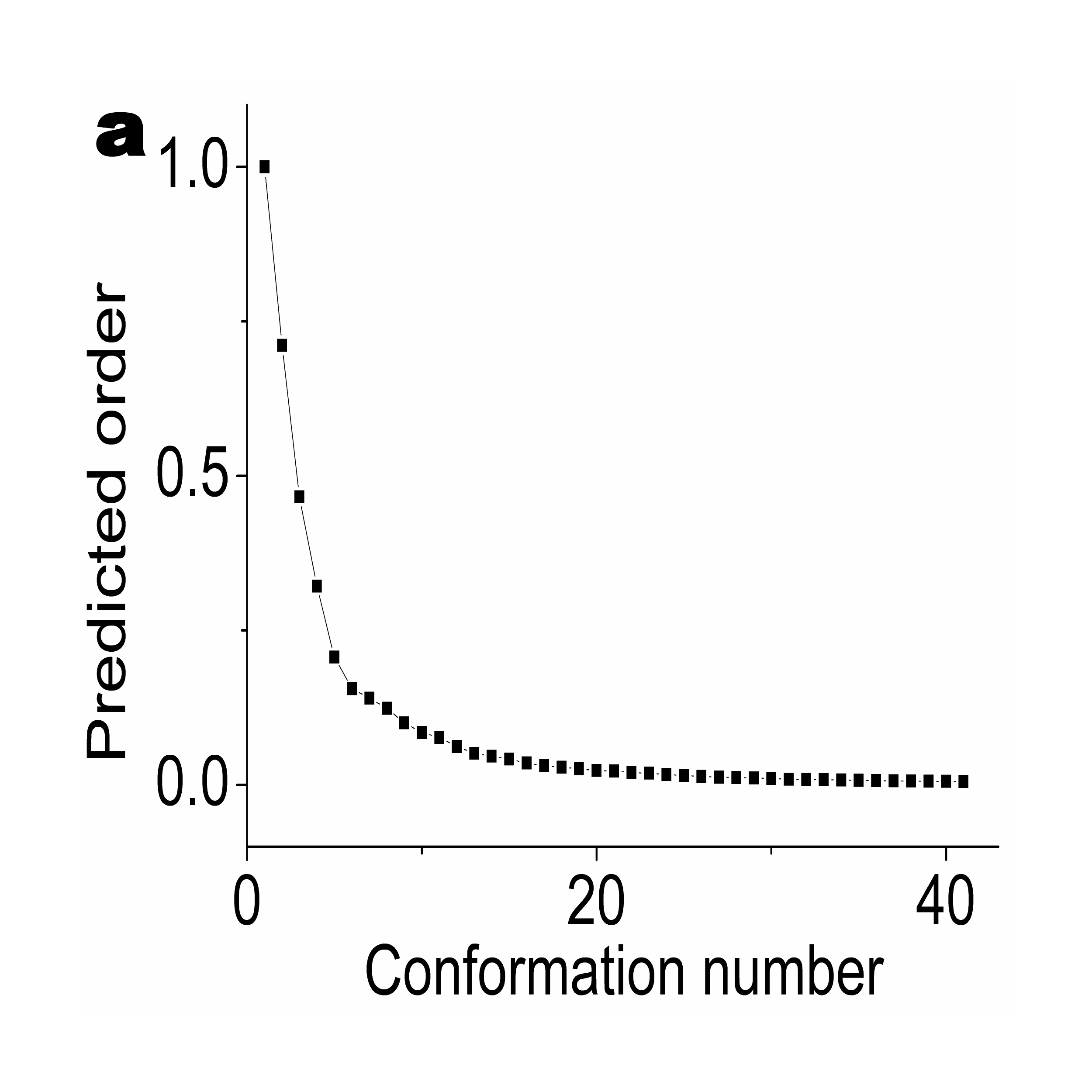} &
\includegraphics[width=0.45\textwidth]{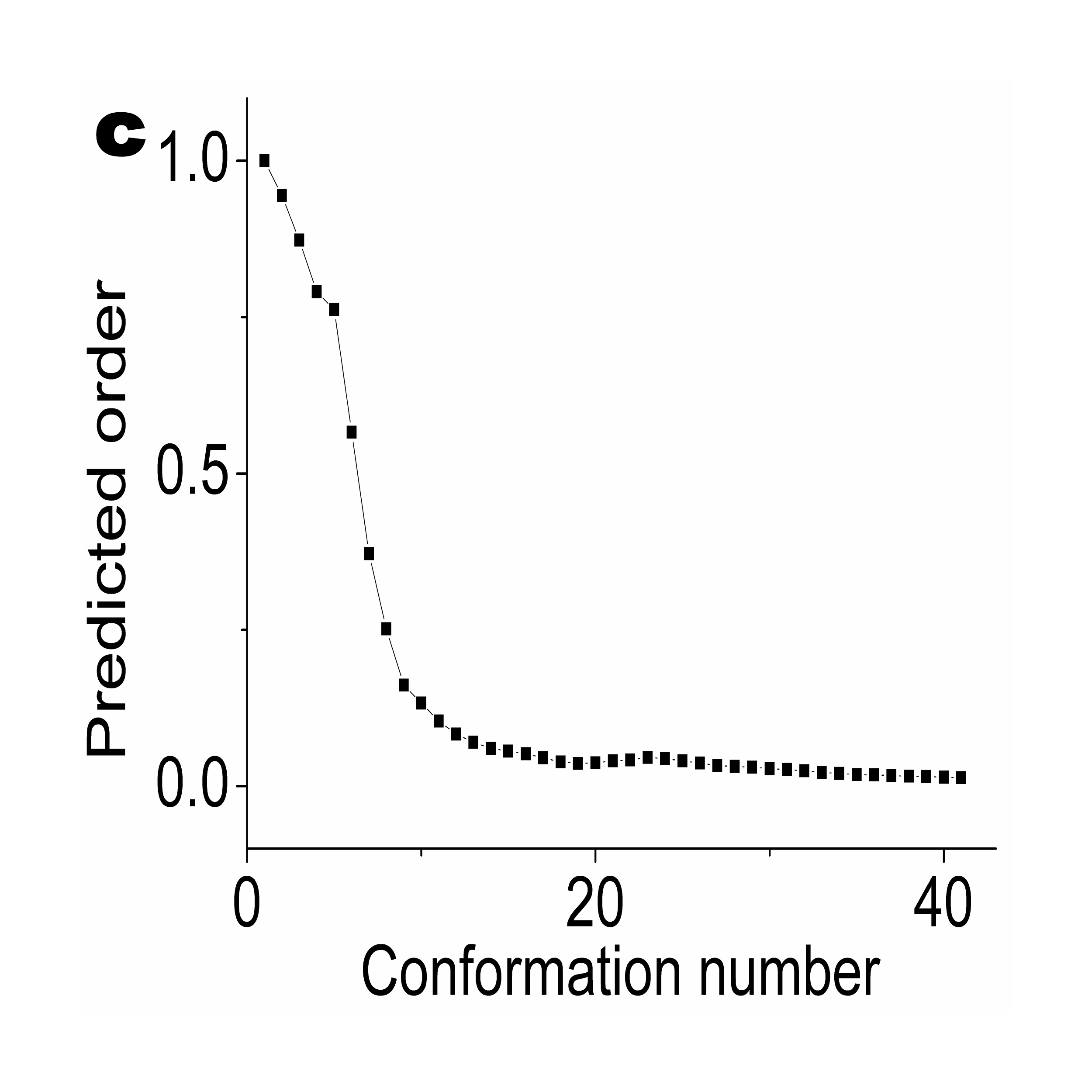} \\
\includegraphics[width=0.45\textwidth]{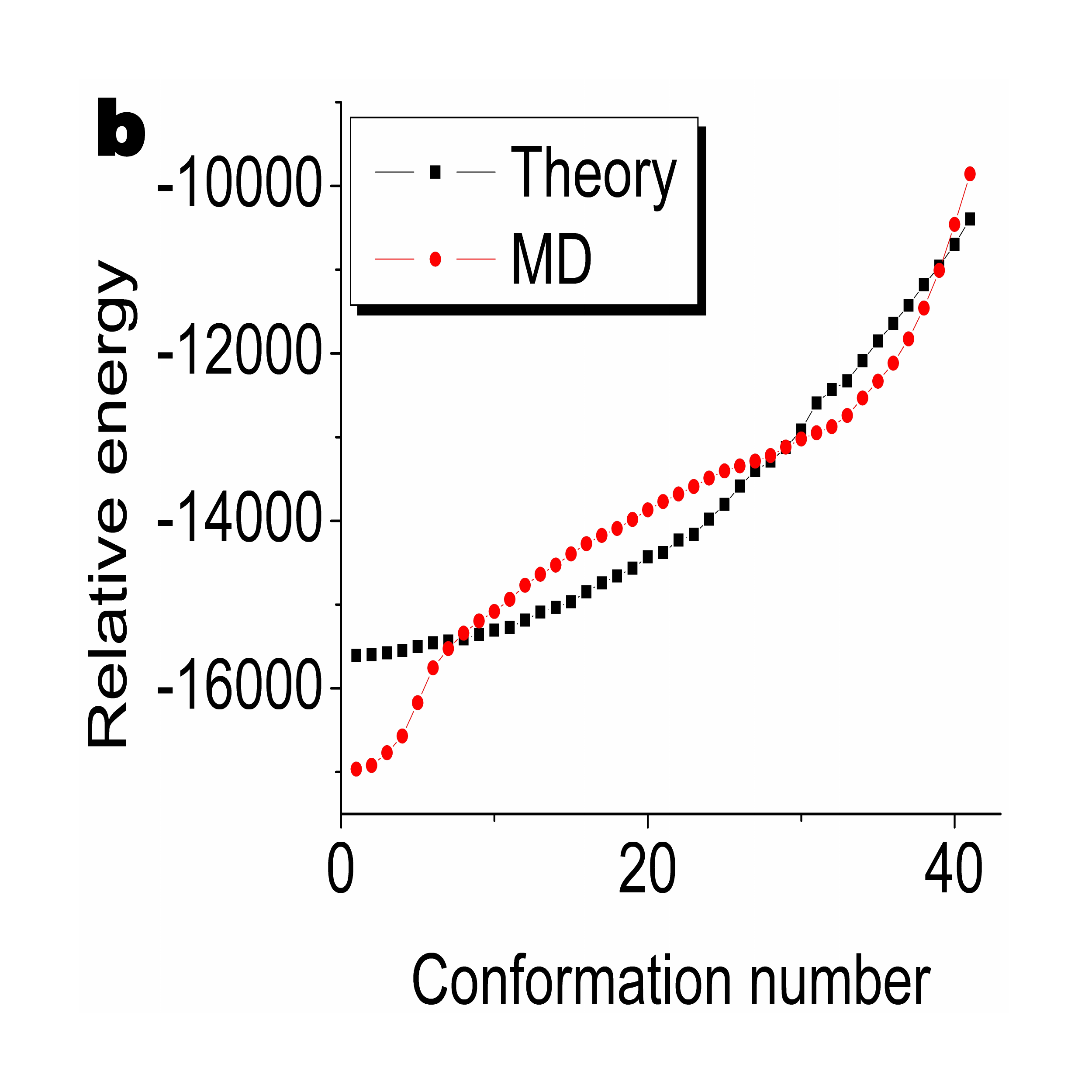} &
\includegraphics[width=0.45\textwidth]{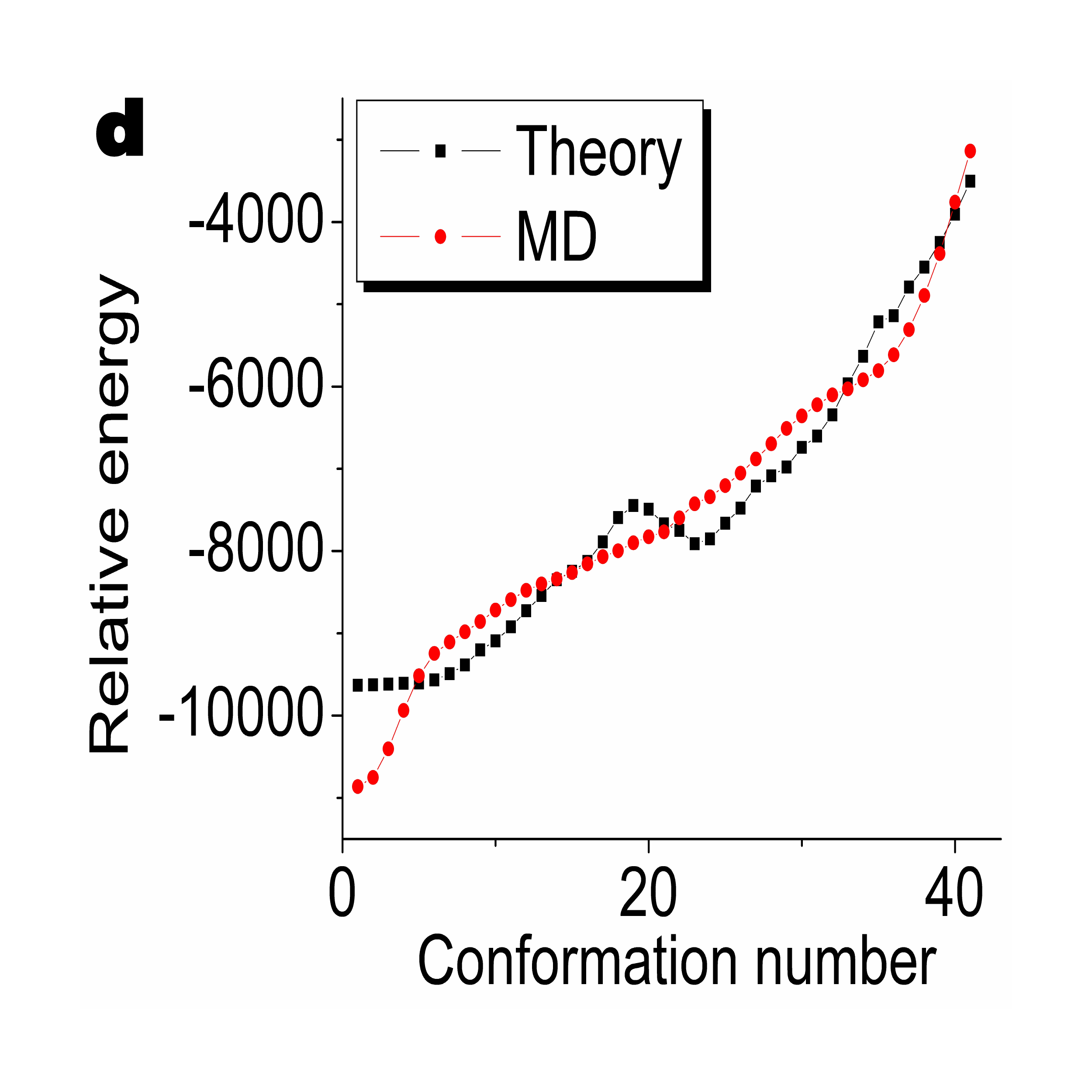}
\end{tabular}
\end{center}
\caption{ The order (top row) and relative energy (bottom row) of conformations for 1ubq (left column) and 2mcm (right column) predicted by using the stability analysis of the ILDM. All conformations are generated from steered molecular dynamics with water molecules. Energies are compared with those obtained from SMD simulations. The correlations between the currently predicted energies and those of SMD simulations are 0.943 and 0.969, respectively for
1ubq and 2mcm.
}
\label{fig:S3}
\end{figure}

\subsubsection{ SMD simulations with water molecules}\label{SMD}

To demonstrate that the present findings are independent of the method used for SMD simulations, we further perform molecular dynamics simulations with a water sphere as described in the literature \cite{Hui:1998}.

In our simulations, we solvate the protein with water molecules represented by the TIP3P model. The smallest possible water sphere which completely immerses the protein is employed. We use the  time step of 2ps in our simulations. The system is minimized for 10000 time steps, and then a total of 15000 time step equilibrium is done with a spherical boundary condition at 310K temperature. The final state is extracted and the SMD simulations are carried out on  protein and water complexes. Basically, we use the constant velocity pulling with spring constant 7 kcal/mol \AA~ and constant velocity of 0.005 \AA. We perform  simulations on two proteins, namely,  1ubq and 2mcm. The total SMD time step is 60000 for 1ubq and 80000 for 2mcm. We extract structures and associated total energies from the simulations.

To show that similar  findings can be obtained from SMD simulations with water molecules, we carry out  ILDM analysis.  Here we use $\sigma_{ij}=4$\AA, $\alpha=10$, $\gamma=28$ and $\beta=\frac{8}{3}$. Figure \ref{fig:S2} plots four typical conformations generated by using the SMD simulations with water molecules for protein 2mcm. Overall, the protein unfolds during the simulation due to the pulling forces. However, compared to  conformations illustrated in Fig. 2 generated by SMD simulations without water molecules,  protein 2mcm clearly shows a different unfolding pathway in the present simulations. It is interesting to know whether a different unfolding pathway changes the theoretical findings from ILDM analysis. Figure \ref{fig:S3} presents the orders and relative energies of two sets of protein conformations analyzed by using the ILDM. Qualitatively,  these results are very much similar to those in Fig.  3 generated by using the  SMD simulations without water molecules.

\begin{figure}
\begin{center}
\begin{tabular}{cc}
\includegraphics[width=0.5\textwidth]{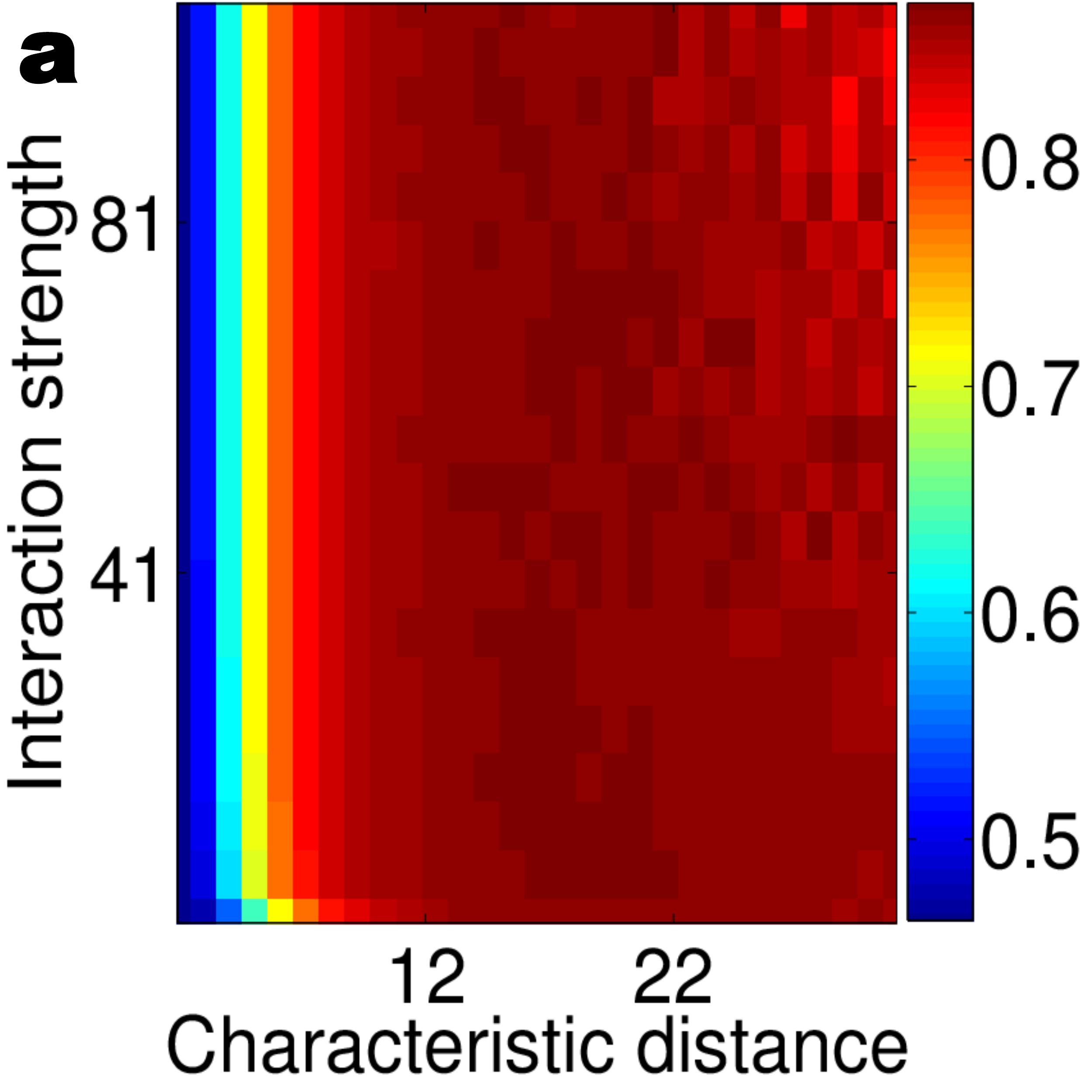}&
\includegraphics[width=0.5\textwidth]{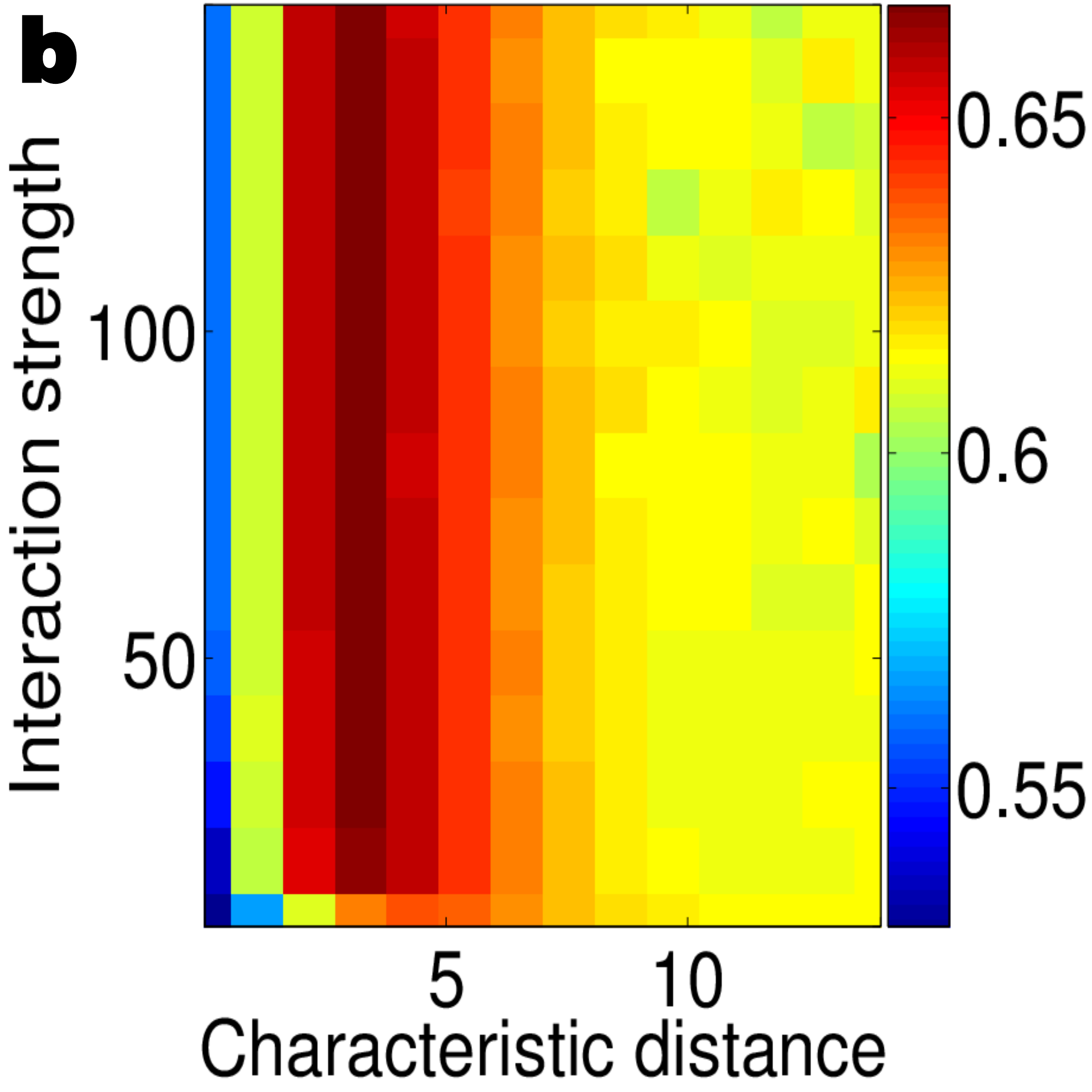}
\end{tabular}
\end{center}
\caption{Phase diagrams of correlation coefficients predicted by the chaotic dynamics model.
{\bf a} Protein 2nuh.  In this case, the performance is not sensitive to characteristic.
{\bf b} Protein 1uku.  In this case,  a narrow range of characteristic length values ($\sigma \approx 5-6$\AA) delivers good correlation coefficients.
}
\label{fig:S4}
\end{figure}

\subsection{Parameters used in the ILDM based  B-factor prediction }

In Fig. 4, we chose the chaotic dynamical system parameters as $\alpha=1$, $\gamma=12$, and $\beta=\frac{8}{3}$.
 Both  forward Euler scheme and  the fourth order Runge-Kutta scheme are used in the time integration to validate each other. Appropriate time increment that satisfies the stability requirement under given interaction strength is used.  Random initial data in the range of [0,1] are used for all oscillators. After the synchronization at $t=t_s$, we set a perturbation on the $j$th particle: $z_j(t_s)\rightarrow 2z_j(t_s)$.  The perturbation procedure is repeatedly carried out for all particles in the protein to compute their relaxation time values, which are converted to B-factors by linear regressions.
For the predictions of B-factors in   Fig. 4, $\sigma_{ij}=20$\AA~ and $\varepsilon=86$ is employed for 2nuh, $\sigma_{ij}=20$\AA~ and $ \varepsilon=21$ are used for 1aru, and  $\sigma_{ij}=5$\AA~ and $ \varepsilon=50$ are chosen for 4dr8.
The above parameter selections appear to be quite random. Indeed, as shown in Fig.  \ref{fig:S4}, a wide range of parameters is admissible for many proteins.

 \begin{table}
\caption{Correlation coefficients $C_c$ of B-factors obtained with the chaotic dynamics model  for 30 proteins ($\varepsilon =50$)}
\begin{center}
\begin{tabular}{c|c| c c||c|c| cc|| c|c|cc}
 \hline
   ID & $N$ & $\sigma=5$\AA &  $\sigma=7$\AA&ID & $N$ & $\sigma=5$\AA   & $\sigma=7$\AA &  ID & $N$ & $\sigma=5$\AA &$\sigma=7$\AA \\\hline
  1AJJ   &37  & 0.646 & 0.645    &1BKF   &107 & 0.525 & 0.537 &1BPI   &58      & 0.570   & 0.542 \\\hline
  1A5W   &146 & 0.759 & 0.755    &1FRD   &98  & 0.646 & 0.577 & 1GK7   &39     & 0.764   & 0.813 \\\hline
  1L11   &164 & 0.544 & 0.574    &1NOA   &113 & 0.615 & 0.600 & 1P9I   &29     & 0.738   & 0.710 \\\hline
  1PMY   &123 & 0.677 & 0.659    &1POA   &118 & 0.645 & 0.589 & 1PZ4   &113    & 0.863   & 0.854 \\\hline
  1QAU   &112 & 0.662 & 0.652    &1A6Q   &363 & 0.618 & 0.638 & 2IHL   &129    & 0.692   & 0.690  \\\hline
  2MCM   &112 & 0.792 & 0.807    &2NLS   &36  & 0.526 & 0.486 & 3APR   &329    & 0.660   & 0.644 \\\hline
  3GRS   &461 & 0.623 & 0.616    &3SEB   &237 & 0.751 & 0.785 & 3VPZ   &322    & 0.540   & 0.542 \\\hline
  3VUB   &101 & 0.622 & 0.620    &3W4Y   &350 & 0.549 & 0.546 & 3ZET   &563    & 0.727   & 0.709 \\\hline
  4AF1   &410 & 0.652 & 0.665    &4AMS   &367 & 0.595 & 0.552 & 4FGY   &276    & 0.738   & 0.769 \\\hline
  4ILR   &279 & 0.673 & 0.706    &4IP3   &343 & 0.467 & 0.447 & 7RSA   &124    & 0.669   & 0.677 \\\hline
\end{tabular}
\label{table:S1}
\end{center}
\end{table}

\begin{table}
\caption{Correlation coefficients $C_c$ of B-factors obtained with Gaussian network model  for 30 proteins}
\begin{center}
\begin{tabular}{cc|cc|cc}
 \hline
   ID \AA & $C_c$& ID\AA   & $C_c$&ID\AA & $C_c$ \\\hline
  1AJJ   &0.718    &1BKF   &0.422 &1BPI   &0.490 \\\hline
  1A5W   &0.690    &1FRD   &0.793 & 1GK7  &0.821 \\\hline
  1L11   &0.584    &1NOA   &0.615 & 1P9I  &0.625 \\\hline
  1PMY   &0.685    &1POA   &0.718 & 1PZ4  &0.843 \\\hline
  1QAU   &0.620    &1A6Q   &0.689 & 2IHL  &0.731  \\\hline
  2MCM   &0.820    &2NLS   &0.530 & 3APR  &0.699 \\\hline
  3GRS   &0.518    &3SEB   &0.827 & 3VPZ  &0.514 \\\hline
  3VUB   &0.607    &3W4Y   &0.444 & 3ZET  &0.683 \\\hline
  4AF1   &0.601    &4AMS   &0.722 & 4FGY  &0.742 \\\hline
  4ILR   &0.695    &4IP3   &0.418 & 7RSA  &0.691 \\\hline
\end{tabular}
\label{table:SGNM1}
\end{center}
\end{table}

\subsection{ ILDM based macromolecular flexibility analysis}

One of practical applications of the ILDM is the flexibility analysis of macromolecules.  The proposed method is based on transverse stability of the ILDM  to predict atomistic B-factors, or temperature factors, of a given molecular structure.  B-factor is a measure of the mean-squared atomic displacement due to thermal motion and possible experimental uncertainties. In general, an atom with a larger B-factor implies it is more flexible and atoms with smaller B-factors are relatively rigid. The analysis of B-factors  provides insights on the large-scale and long-time functional behaviors of native state macromolecules.  This information  is complementary to that obtained from atomic detail simulation techniques.

There are many other interesting approaches for the flexibility analysis in the literature.  For example, normal mode analysis (NMA)  has been proposed to uncover the intrinsic structural flexibility of the protein \cite{Go:1983,Tasumi:1982,Brooks:1983,Levitt:1985}, and  study  biomolecular systems like  lysozyme\cite{Levitt:1985,Brooks:1985}. NMA approach has a large number of variations. Tirion proposed elastic network model (ENM) by simplifying the interaction potential in the NMA  \cite{Tirion:1996}. By introducing the idea from polymer science \cite{Flory:1976}, Bahar {\it et al.}  use Gaussian network model (GNM) to describe  protein flexibility \cite{Bahar:1997,Bahar:1998}. They employed  the C$_\alpha$ atom  representation of proteins  based on local packing density and contact topology. Anisotropic fluctuations are considered in anisotropic network model (ANM) \cite{Atilgan:2001}. Parameters in these models are calibrated with Debye-Wallers factor or B-factors. Crystal structures have been also taken into considerations \cite{Kundu:2002,Kondrashov:2007,Hisen:2008,GSong:2007}. Due to the simplified potential and reduced representation, these coarse-grained based ENM and GNM approaches \cite{Bahar:1997,Bahar:1998,Atilgan:2001,Hinsen:1998,Tama:2001,LiGH:2002} become popularity and have been applied to the study of macroproteins or protein complexes, such as, hemoglobin \cite{CXu:2003}, F1 ATPase \cite{WZheng:2003,QCui:2004}, chaperonin GroEL \cite{Keskin:2002,WZheng:2007}, viral capsids \cite{Rader:2005,Tama:2005}, and ribosome \cite{Tama:2003,YWang:2004}. More applications can be found in a few good review papers \cite{JMa:2005,LWYang:2008,Skjaven:2009,QCui:2010}. Our ILDM based method should be potentially useful for solving these problems as well.

\subsubsection{The robustness  of ILDM based  protein flexibility analysis  }

The performance of the present chaotic dynamics model for protein flexibility analysis depends on two parameters, namely, the characteristic length $\sigma$ and the overall interaction strength $\varepsilon$. It is important to know the parameter region that delivers satisfactory predictions. The performance of a theoretical prediction of B-factors is typical measured by the
correlation coefficient
\begin{eqnarray}\label{correlation}
   C_c=\frac{\Sigma^N_{i=1}(B^e_i-<B^e> )( B^t_i-<B^t> )}
   {\sqrt{\Sigma^N_{i=1}(B^e_i-<B^e>)^2\Sigma^N_{i=1}(B^t_i-<B^t>)^2}},
\end{eqnarray}
where $\{B^t_i,  i=1,2,\cdots,N\}$ are a set of B-factors predicted by theoretical calculations and $\{B^e_i, i=1,2,\cdots, N\}$ are a set of B-factors obtained from experiment measurements, such as  X-ray  scattering or  neutron scattering. Here $<B^t>$ and $<B^e>$ are the statistical averages of theoretical and experimental B-factors, respectively.

\begin{figure}
\begin{center}
\begin{tabular}{ccc}
\includegraphics[width=0.32\textwidth]{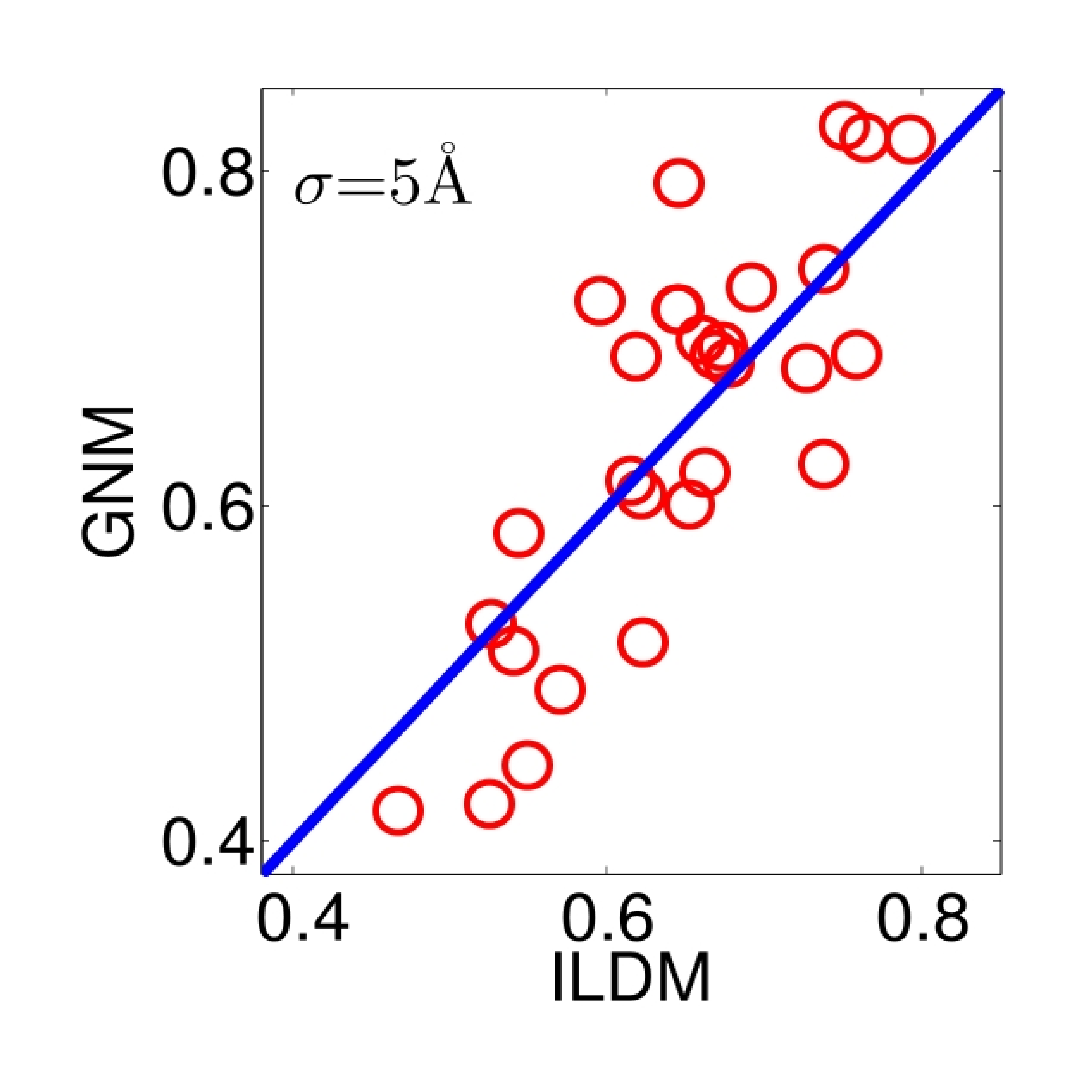} &
\includegraphics[width=0.32\textwidth]{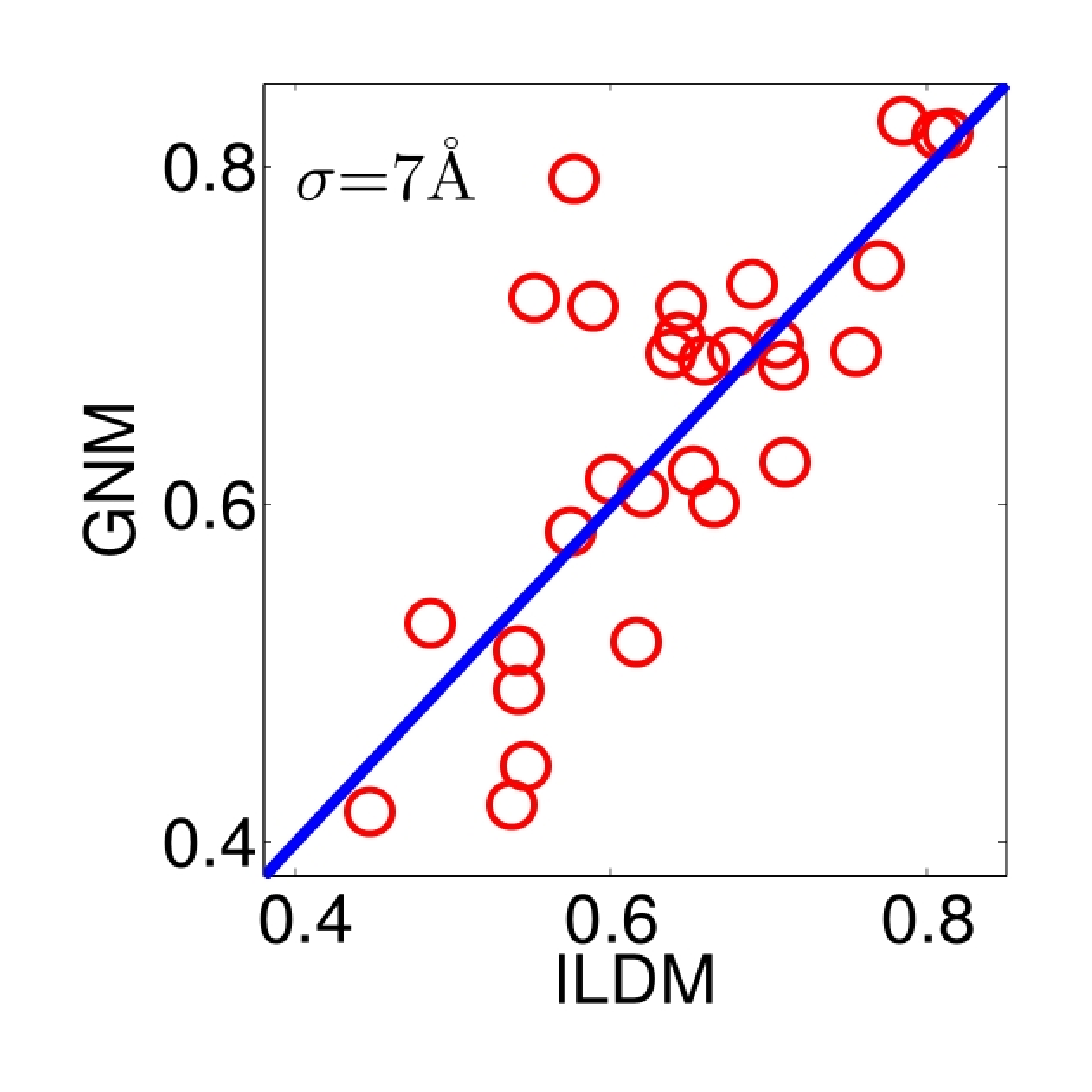} &
\includegraphics[width=0.32\textwidth]{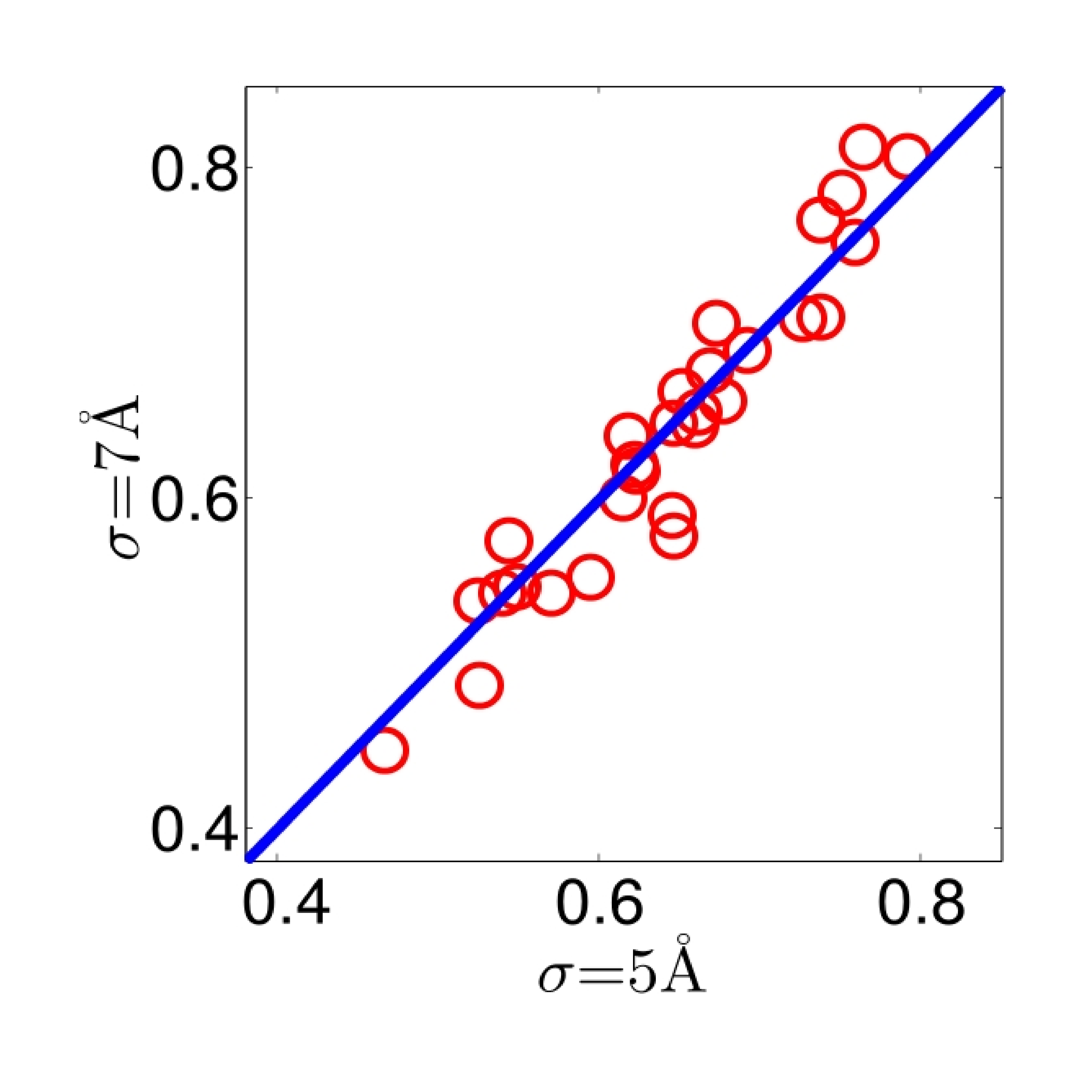}
\end{tabular}
\end{center}
\caption{ The comparison of the B-factor predicted by ILDM and GNM in terms of correlation coefficients.
{\bf a} The Comparison of the B-factor predicted by ILDM with $\sigma=5$\AA ~and GNM.
{\bf b} The Comparison of the B-factor predicted by ILDM with $\sigma=7$\AA ~and GNM.
{\bf c} The Comparison of the B-factor predicted by ILDM with $\sigma=5$\AA ~and $\sigma=7$\AA.
 }
\label{fig:S5}
\end{figure}

Figure \ref{fig:S4} demonstrates the correlation coefficients of the present method for two proteins  2nuh and 1uku. It is seen that the performance of the proposed method is not sensitive to the interaction strength $\varepsilon$ -- there is a wide range  of $\varepsilon$ values that deliver good B-factor predictions.  However, the sensitivity of the characteristic length may vary from protein to protein. For some proteins,  a wide range of $\sigma$ values can be used, see  Figure \ref{fig:S4}{\bf a}. However, for many other proteins, only a narrow range of   $\sigma$ values works well, as shown in Figure \ref{fig:S4}{\bf b}.

To further demonstrate the robustness of the proposed chaotic dynamics method for protein flexibility analysis, we have computed the B- factors of a large number of proteins with diversified sizes.  Since Fig. \ref{fig:S4} indicates that the present chaotic dynamics model is not sensitive to the interaction strength,   the same interaction strength value, $\varepsilon =50$, is employed in all numerical calculation. Two characteristic lengths, $\sigma=5$\AA~ and $\sigma=7$\AA, are examined. ILDM based correlation coefficients are summarized in  Table  \ref{table:S1}, where $N$ is the number of residues. It is seen that our results are not very sensitive to characteristic lengths. Typically, good correlation coefficients can be attained for $\sigma \approx 5-10$\AA.  The related results from Gaussian network model (GNM) \cite{JKPark:2013} are listed in Table \ref{table:SGNM1}.

\begin{figure}
\begin{center}
\begin{tabular}{cc}
\includegraphics[width=0.45\textwidth]{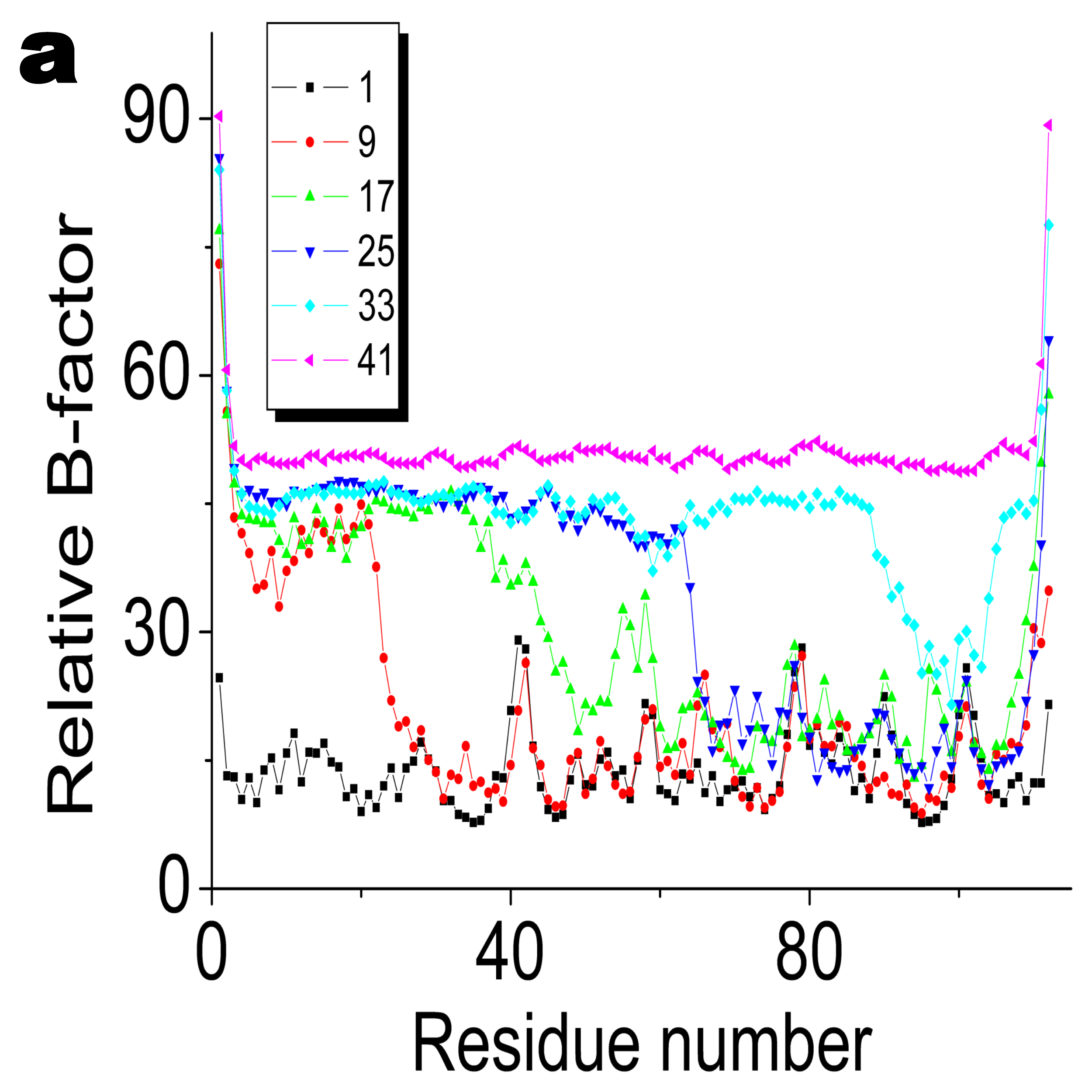}
\includegraphics[width=0.45\textwidth]{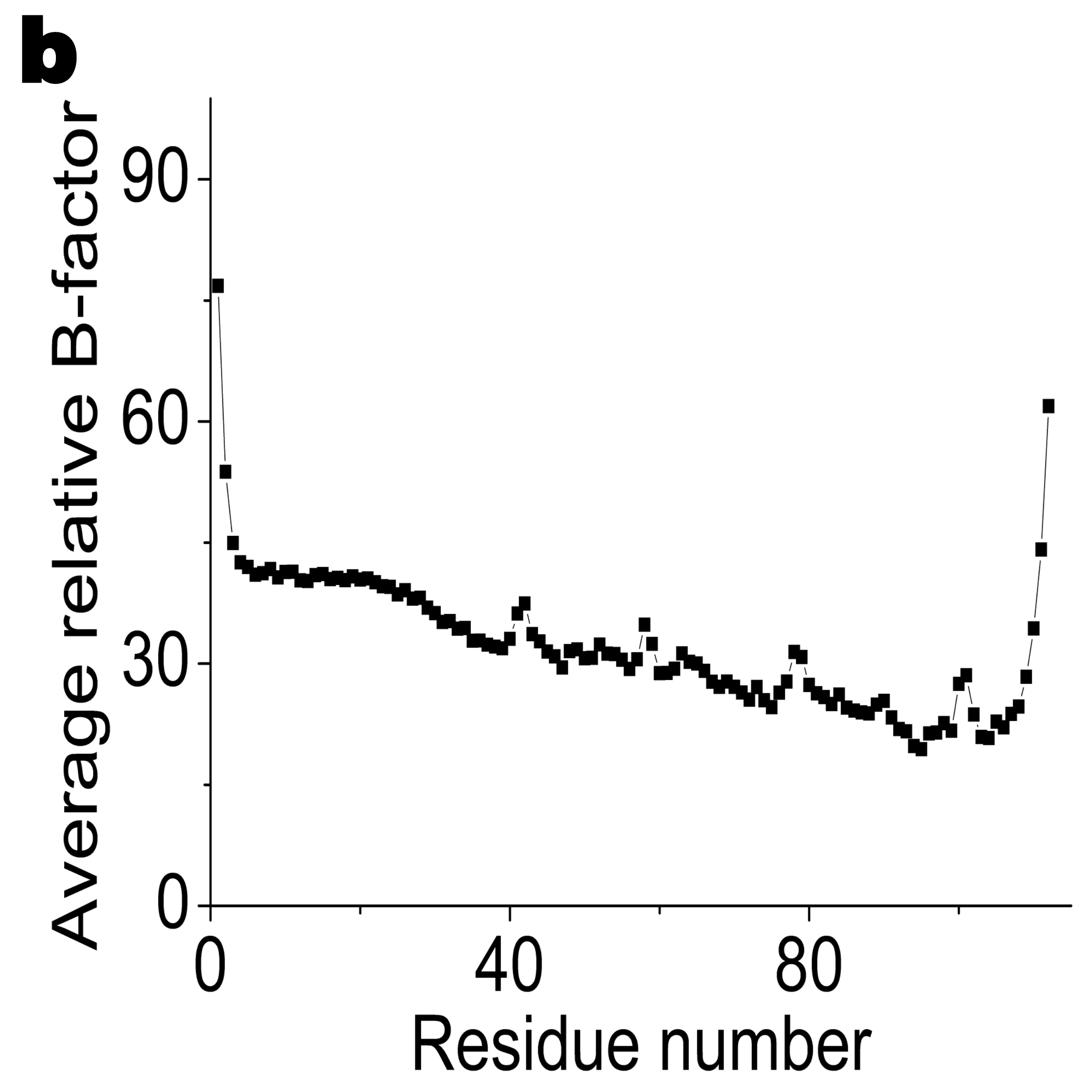}
\end{tabular}
\end{center}
\caption{  The relative B-factors of protein 2mcm predicted by the chaotic dynamics model. All partially folded and completely unfolded  conformations are  generated by a constant pulling force.
{\bf a} The relative B-factors  for  6 conformations. Clearly residue numbers 90 to 99 show a strong resistance to the pulling force compared to  other residues.
{\bf b} The relative B-factors averaged over all of 41 conformations.  The averaged relative B-factors show the persistence of residue interactions.
}
\label{fig:S6}
\end{figure}

Figure \ref{fig:S5} shows the performance of  the proposed ILDM method for B-factor prediction in a comparison with the state of the art GNM approach. The cutoff parameter of 7\AA~ is used for the GNM. It is clear from Figs. \ref{fig:S5}{\bf a} and  \ref{fig:S5}{\bf b}  that the proposed ILDM method does an excellent job relative to the GNM. The consistency in Fig.  \ref{fig:S5}{\bf c} indicates that the present method is not very sensitive to the parameter.

\subsubsection{Relative flexibility of residues}

The proposed chaotic dynamics model   can be used to analyzed the relative rigidity of residues under a constant pulling force. This analysis goes beyond the harmonicity and elasticity assumption used in most other flexibility approaches.  In Fig.  \ref{fig:S6}{\bf a}, we plot B-factors of 6 conformations generated by pulling protein structure 2mcm. It is seen that residues at both  C terminal and N terminal are very sensitive to pulling forces. Clearly  residues 90-99 have smaller relative B-factors in all conformations and thus are more rigid than others.   Figure  \ref{fig:S6}{\bf b} plots relative B-factors averaged over 41 conformations. Averaged relative B-factors reveal the dynamical stability of protein residues  under external pulling forces, whereas the B-factors of the native structure indicate the static rigidity of protein residues.

\begin{figure}
\begin{center}
\begin{tabular}{cc}
\includegraphics[width=0.47\textwidth]{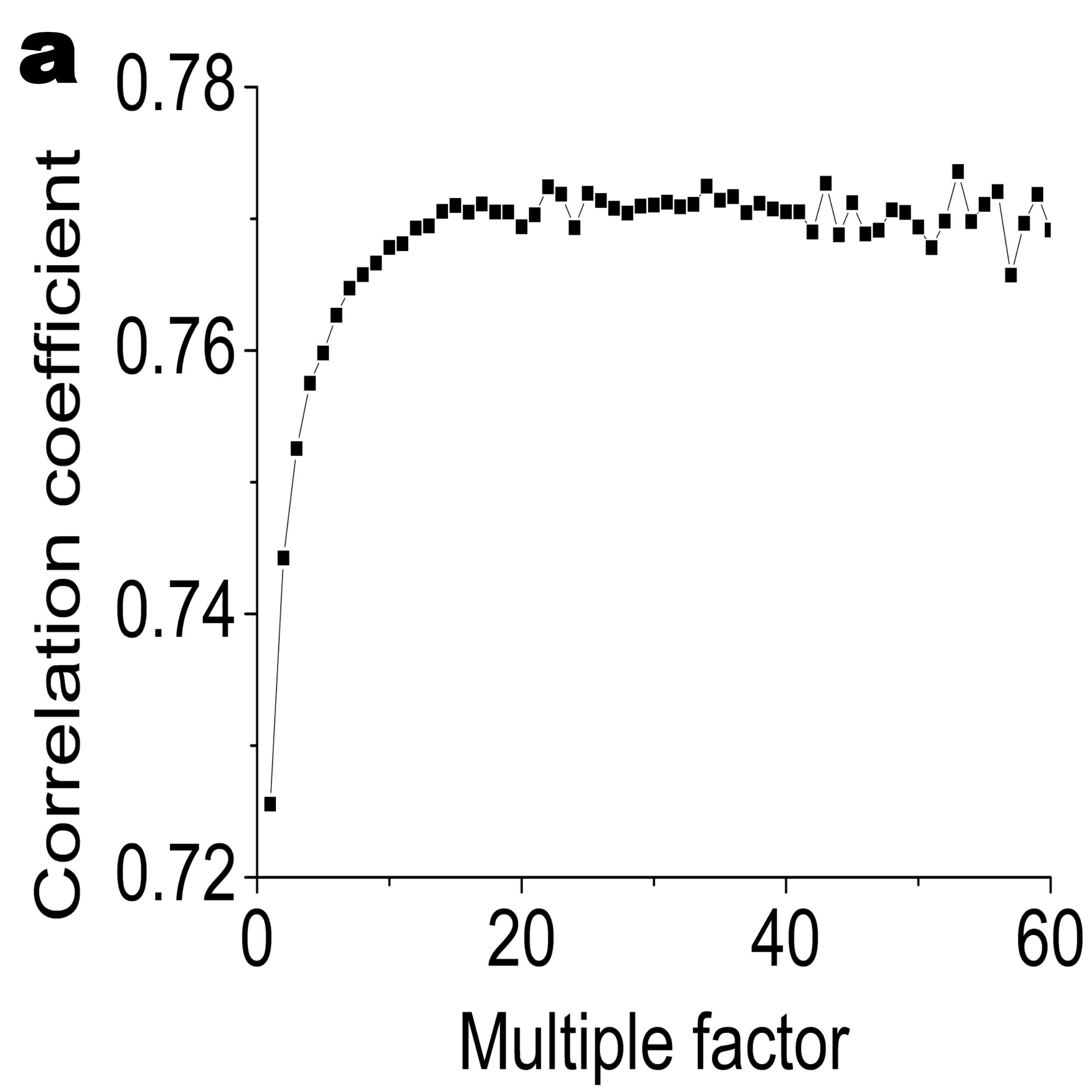}&
\includegraphics[width=0.47\textwidth]{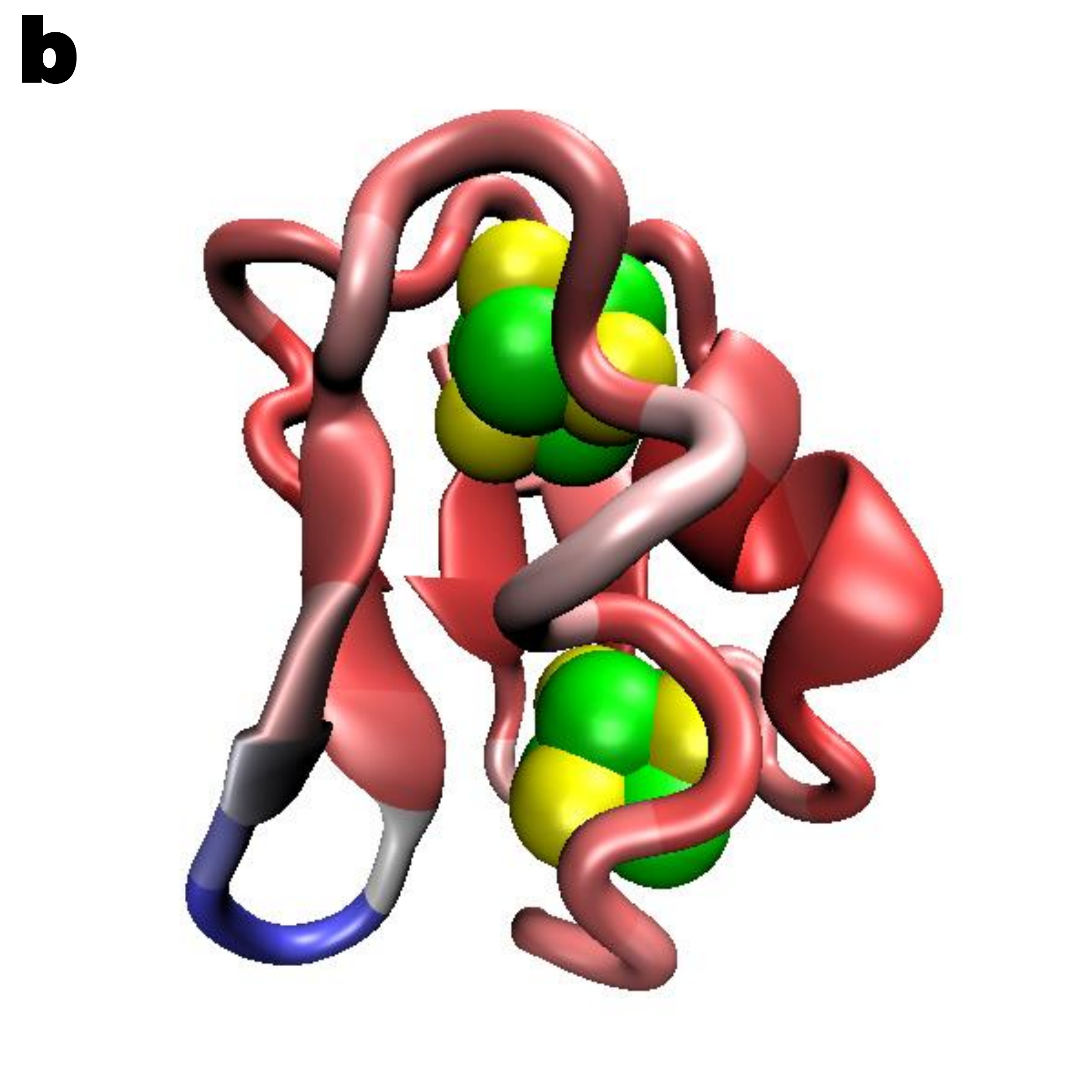}
\end{tabular}
\end{center}
\caption{ The improvement of B-factor prediction by considering cofactors.
{\bf a} The impact  of the interaction strength  of two Fe$_4$S$_4$ clusters in 1fca to the predicted correlation coefficient.
{\bf b} The structure of ifca showing two Fe$_4$S$_4$ clusters.  Amino acid residues are colored according to B-factor values.
}
\label{fig:S7}
\end{figure}

\subsubsection{Improvement of the flexibility prediction  by  considering  co-factors}

Protein structures downloaded  from the PBD typically contain many cofactors, i.e., coenzymes and prosthetic groups,  which are important for proteins' biological functions. Cofactors also contribute to protein structural rigidity. Therefore, the inclusion of cofactors in the present chaotic dynamics model will improve the prediction of protein B-factor. In the present work, we consider a simple treatment of cofactors in which the nonlinear oscillators of cofactors are used as part  of driven sources while those of residues are treated as a response system. This treatment saves computational cost if the MND is on a stable fixed point, because one does not need to actually compute  cofactors'  dynamics.

 Figure \ref{fig:S7} depicts the improvement in the theoretical prediction of protein 1fac B-factors due to  consideration of two metal clusters, i.e., Fe$_4$S$_4$ clusters. We plot the interaction strength of cofactors with respect to the correlation coefficient. Clearly, the consideration of cofactors leads to a five percent improvement in our prediction.

\section*{Acknowledgments}

This work was supported in part by NSF grants  DMS-1160352 and IIS-1302285, and NIH Grant R01GM-090208.
The authors acknowledge the Mathematical Biosciences Institute for hosting valuable workshops.

\vspace{0.6cm}
\clearpage


\end{document}